\documentclass[structabstract]{aa}  
\usepackage{graphicx}
\usepackage{color}
\usepackage{colortbl}
\usepackage{amssymb}
\usepackage{amsmath}
\usepackage{longtable}
\usepackage{txfonts}
\usepackage{multirow}
\usepackage{textcomp}
\usepackage{natbib}
\bibpunct{(}{)}{;}{a}{}{,} 
\begin{document}

\title{Properties of star forming galaxies in AKARI Deep Field-South\thanks{Table A.1 is only available in electronic form at the CDS via anonymous ftp to cdsarc.u-strasbg.fr (130.79.128.5) or via http://cdsweb.u-strasbg.fr/cgi-bin/qcat?J/A+A}}

 \author{K.\ Ma{\l}ek\thanks{Postdoctoral Fellow of the Japan Society for the Promotion of Science, \email{malek.kasia@nagoya-u.jp}}\inst{1}, A.\ Pollo \inst{2,3}, T.\ T.\ Takeuchi \inst{1},  V.~Buat \inst{4}, D.~Burgarella \inst{4},  M.~Malkan \inst{5}, E.~Giovannoli\inst{6}, A.~Kurek\inst{3}, S.~Matsuura\inst{7}  }

\institute{
	Department of Particle and Astrophysical Science, Nagoya University, Furo-cho, Chikusa-ku, 464-8602 Nagoya, Japan
	\and National Centre for Nuclear Research, ul. Ho\.za 69, 00-681 Warszawa, Poland  
	\and The Astronomical Observatory of the Jagiellonian University, ul.\ Orla 171, 30-244 Krak\'{o}w, Poland 
	\and Laboratoire d'Astrophysique de Marseille, OAMP, Universit$\rm{\acute{e}}$ Aix-Marseille, CNRS, 38 rue Fr$\rm{\acute{e}}$d$\rm{\acute{e}}$ric Joliot-Curie, 13388 Marseille, cedex 13, France
        \and Department of Physics and Astronomy, University of California, Los Angeles, CA 90095-1547
	\and University of the Western Cape, Private Bag X17, 7535, Bellville, Cape Town, South Africa
	\and Institute of Space and Astronautical Science, Japan Aerospace Exploration Agency, Sagamihara, 229-8510 Kanagawa, Japan}
\authorrunning{K. Ma{\l}ek et al.,}
 
   \date{Received xxx; accepted xxx}
 
  \abstract
   {}
   {The main aim of this work is the characterization of physical properties  of galaxies detected in the far infrared (FIR) in the AKARI Deep Field-South 
(ADF--S) survey.}
   {Starting from a catalog of the  1\,000 brightest ADF--S sources in the WIDE--S (90$\mu$m) AKARI band, we constructed a subsample of galaxies with spectral coverage from the ultraviolet to the far infrared. 
We then analyzed the multiwavelength properties of this 90$\mu$m selected sample of galaxies.
For galaxies without known spectroscopic redshifts we computed photometric redshifts using the codes \textit{Photometric Analysis for Redshift Estimate} (Le~PHARE) and \textit{Code Investigating GALaxy Emission} (CIGALE), tested these photometric redshifts using spectroscopic redshifts, and compared the performances of both codes.
To test the reliability of parameters obtained by fitting spectral energy distributions, a mock cataloge was generated.   
 }
   {We built a large multiwavelength catalog of more than 500 ADF--S galaxies.  
We successfully fitted spectral energy distributions of 186 galaxies with $\rm{\chi^2_{min}<4}$, and analyzed the output parameters of the fits.
We conclude that our sample consists mostly of nearby actively star-forming galaxies, and all our galaxies have a relatively high metallicity.
We estimated photometric redshifts for 113 galaxies from the whole ADF--S sample. 
Comparing the performance of Le~PHARE and CIGALE, we found that CIGALE gives more reliable redshift estimates for our galaxies, which implies that including the IR photometry allows for substantial improvement of photometric 
redshift estimation.

}
   {}

   \keywords{surveys -- galaxies: fundamental parameters, star formation, evolution -- infrared: galaxies}

   \maketitle

\section{Introduction}

Recent advances in satellite observations have extended our knowledge of galaxies and their evolution  by measuring their properties at  long wavelengths.
In particular the far IR (FIR, $\mathrm{\lambda > 100 \mu m}$)  is strongly linked with star-formation activity (SF) in galaxies which can be studied across cosmic history  \citep{decunha10}. 
Since star-forming regions are dust-enshrouded in the dense cores of molecular clouds,  the earliest stages can be observed at mm wavelengths. 
When the clouds collapse, and the proto-stars form, the dust near them starts emitting in the near and mid-infrared range. 
In the next step of star formation, the warmest regions of the cloud, around the newly formed young stars, are heated by their UV emission, and this energy is re-radiated in the  far infrared. 

The first all-sky IR cataloge was produced by the Infra-Red Astronomy Satellite (IRAS) \citep{neugebauer84, beichman87}.
The Japanese satellite AKARI dedicated to infrared (IR) astronomy \citep{murakami07}, provided second-generation infrared cataloges.
The primary purpose of the AKARI mission was to obtain a cataloge with better spatial resolution and wider spectral coverage than IRAS. 
To improve the knowledge of IR astrophysical sources, in addition to an all-sky survey,  AKARI provided deeper observations of two fields centered on the north and south ecliptic poles \citep{wada08,takagi12, matsura11}.  

In this article we analyze data from one of these fields, located near the south ecliptic pole, referred to as the AKARI Deep Field South (ADF--S). 
This field has the lowest Galactic cirrus emission in the whole sky, i.e., lower than 0.5 MJy$\rm{sr^{-1}}$ at 100$\mu$m \citep{schleger98}.
This region allows for the best FIR extragalactic image of the Universe.
The on-board AKARI instrument called  the Far-Infrared Surveyor (FIS: \citealp{kawada07}), produced a FIR map which covers approximately 12 square degrees, centered at 
$\mathrm{RA}=4^{\rm{h}}44^{\rm{m}}00^{\rm{s}}$, $\mathrm{DEC=-53\degr20\arcmin00\farcs0}$  J2000 \citep{shirahata09b}.
This survey was done in four photometric bands: 65$\mu$m, 90$\mu$m, 140$\mu$m, and 160$\mu$m; 2\,263 infrared sources were detected down to $\sim$ 20 mJy in the 90$\mu$m band. 

The first analysis of this sample in terms of nature and properties, for 1\,000 ADF--S objects brighter than 0.0301~Jy (which corresponds to the 6$\sigma$ detection limit) in the 90$\mu$m band, was presented in \cite{malek10}. 
The ADF--S field was also observed in other wavelength ranges, i.e., at millimeter and submillimeter wavelengths \citep{hatsukade11}, radio wavelengths \citep{white12}, and mid- and far-infrared wavelengths \citep{clements11}.  
Additionally, dedicated spectroscopic measurements of selected ADF--S sources were performed by \cite{sedgwick11} in order to build the FIR luminosity function of local star-forming galaxies. 

In this article, we present a multiwavelength study of a sample of 545 ADF--S sources identified as galaxies.
Based on the identification of ADF--S sources from \citet{malek10}, and with the addition of new measurements and redshifts (spectroscopic and photometric) obtained subsequently, we analyzed the properties of FIR-bright galaxies. 
For this purpose, we used CIGALE (Code Investigating GALaxy Emission\footnote{http://cigale.oamp.fr}, \citealp{noll09}), which provides physical information about galaxies by fitting spectral energy distributions (SEDs) covering wavelengths from the UV to FIR.
Our main aim is to study this galaxy sample to determine the main characteristics of their star formation and dust emission.

The paper is organized as follows.
In Sect.~\ref{data} we present the data used for our analysis. 
Section~\ref{zsection} describes the spectroscopic redshifts in the sample. 
In section~\ref{sedsection} we present the main input parameters used by CIGALE \citep{noll09} to fit the SEDs of our galaxies, and in section~\ref{selection} we discuss selection criteria used for our analysis.
In the same section we also present tests of reliability of the parameters estimated by CIGALE. 
Basic properties of a sample of galaxies with known spectroscopic redshifts are shown in Sect.~\ref{resultssectionSPECT}.  
Section~\ref{photo_subsection} presents tests of photometric redshift estimation and redshift properties of the whole sample. 
Average SEDs for different types of galaxies are presented in Sect.~\ref{resultAVSEDS}.
Discussion of physical and statistical properties of the obtained SEDs is presented in  Section~\ref{moresection}. 
We conclude in Section~\ref{conclsection}. 

We assume that  $H_{0}$ = 70  [km/s/Mpc],  $\Omega_{m,0}$ =  0.3, and  $\Omega_{\Lambda,0}$ = 0.7.

\section{Data}
\label{data}
The main aim of our work is to build a galaxy sample with high-quality fluxes from the UV to the FIR.
As the starting point, we use the ADF--S multiwavelength cataloge \citep{malek10}, based on 1\,000 ADF--S sources {which are the} brightest in the WIDE--S (90~$\mu$m) AKARI band. 
This cataloge was created by cross-identification of the ADF--S point source catalog (based on $90~\mu$m) with publicly available databases, mainly the SIMBAD\footnote{http://simbad.u-strasbg.fr/simbad/} and the NASA/IPAC Extragalactic Database\footnote{http://nedwww.ipac.caltech.edu/} (NED).
The search for ADF--S counterparts was performed within the radius of 40'' around each source which corresponds to the { apperture} size in two FIS AKARI bands, N60  and WIDE--S, centered on 65 and 90~$\mu$m, respectively \citep{shirahata08}. 
Most of the ADF--S counterparts are located at an angular distance from the source that is significantly smaller than the nominal resolution of the FIS detector ($\sim$7''). 
A few counterparts that are found at larger angular distances from the source are more likely chance coincidences.
The angular distribution of the final sample used for the analysis presented in this paper is discussed in Sect.~\ref{resultssectionSPECT} (for the spectroscopic redshift sample) and Sect.~\ref{photo_subsection} (for the photometric redshift sample).

{ We have also checked angular sizes of the counterparts for all possible sources. 
We have found that for 10\% of our sample the major axis of the optical counterpart was larger than the radius of the aperture used in the WIDE--S AKARI band, and among them, 18 galaxies (4\% of the total galaxy sample) have angular sizes a few or even tens times larger than the aperture radius of the AKARI WIDE--S band. 
We decided to use these data for the subsequent analysis, but - as will be discussed below - all the sources with angular sizes significantly exceeding aperture size were finally rejected from the final sample because of the poor quality of their fitted SEDs.  
Some of the ADF--S sources have possible multiple counterparts; however, in the final sample we left only galaxies for which the visual inspection confirmed that their identifications are most likely the correct ones, and the contamination from other sources can be neglected.
}

Among these, the sample of sources which we could successfully cross-identify with previously known extragalactic objects (observed by other missions)  contains 545 galaxies. 
In addition to the data presented in \citet{malek10}, additional measurements, mostly from Wide-field Infrared Survey Explorer (WISE, \citep{wright10}) and Galaxy Evolution Explorer (GALEX, \citep{dale07}), as well as additional information from public databases (mainly the SIMBAD, NED\, and NASA/IPAC Infrared Science Archive\footnote{http://irsa.ipac.caltech.edu/} - IRSA) were used in our analysis. 
{In $\sim$ 15\% of cases Spitzer/MIPS measurements were taken from the  NED database \citep{scott10}; the remaining Spitzer/MIPS measurements used for our analysis were made directly from images.}
In Table~\ref{measurements_table_1}, we summarize the main data used in the paper.

We have also checked the angular sizes of the counterparts for all possible sources. 
We have found that for 10\% of our sample the major axis of the optical counterpart was larger than the radius of the aperture used in the WIDE--S AKARI band, and among them, 18 galaxies (4\% of the total galaxy sample) have angular sizes a few or even tens of times larger than the aperture radius of the AKARI WIDE--S band. 
We decided to use these data for the subsequent analysis, but - as will be discussed below - all the sources with angular sizes significantly exceeding aperture size were finally rejected from the final sample because of the poor quality of their fitted SEDs.  
Some of ADF--S sources have possible multiple counterparts; however, in the final sample we left only galaxies for which the visual inspection confirmed that that their identifications are most likely the correct ones, and the contamination from other sources can be neglected.

\begin{table}[t]
\caption[]{The main data used for the SED fitting and redshift estimation. 
We have listed the name of the survey, band name and effective wavelength, and number of sources correlated with the ADF--S database. 
$^a$ - \cite{maddox90}, data taken from the NED database. 
Data marked as\textit{ SIMBAD}  were taken from the SIMBAD database: 
$^{(b)}$ - mostly from Dressler's catalog of galaxies in clusters \citep{dressler80b}; a part of the used measurements were taken from the 6dF galaxy survey \citep{jones04}, and the ESO-Uppsala cataloge of galaxies \citep{lauberts89};
$^{(c)}$ - \cite{jones04} or \cite{lauberts89};  
$^{(d,e,f)}$ - 2MASS measurements  \citep{skrutskie06} taken from the SIMBAD database without any additional corrections.}  
\label{measurements_table_1}
\begin{tabular}{l|l |l |l }
Survey	& Band 		& Wavelength 	& $No._{all}$    \\  \hline \hline
GALEX 	& FUV 		& 0.1$\mu$m 	& 78		\\
GALEX 	& NUV 		& 0.2$\mu$m 	& 98		\\
{ APM}$^a$ & $\rm{B_{j}}$ & 0.4$\mu$m 	& 401		\\
SIMBAD$^{(b)}$ 	& B 	& 0.4$\mu$m 	& 55		\\
SIMBAD$^{(c)}$ 	& R 	& 0.6$\mu$m 	& 30 		\\
SIMBAD$^{(d)}$ 	& J 	& 1.2$\mu$m 	& 66 		\\
2MASS 	& J 		& 1.2$\mu$m 	& 204 		\\
SIMBAD$^{(e)}$ 	& H 	& 1.7$\mu$m 	& 66 		\\
2MASS 	& H 		& 1.7$\mu$m 	& 204 		\\
SIMBAD$^{(f)}$ 	& K 	& 2.2$\mu$m 	& 66 		\\
2MASS 	& Ks 		& 2.2$\mu$m 	& 204 		\\
WISE 	& W1 		& 3.4$\mu$m 	& 280		\\
WISE 	& W2 		& 4.6$\mu$m	& 280		\\
WISE 	& W3 		& 12$\mu$m	& 280		\\
IRAS 	& Band-1 	& 12$\mu$m 	& 18 		\\
MIPS 	& 24$\mu$m 	& 23.7$\mu$m 	& 100 		\\
IRAS 	& Band-2 	& 25$\mu$m 	& 19 		\\
IRAS 	& Band-3 	& 60$\mu$m 	& 35		\\
AKARI 	& N60 		& 65$\mu$m 	& 116		\\
MIPS 	& 70$\mu$m 	& 71.4$\mu$m 	& 15 		\\
AKARI 	& WIDE--S 	& 90$\mu$m 	& { 545}		\\
IRAS 	& Band-4 	& 100$\mu$m 	& 30 		\\
AKARI 	& WIDE--L 	& 140$\mu$m 	& 65		\\
AKARI 	& N160 		& 160$\mu$m 	& 9		\\

\end{tabular} 

\end{table}
 
The technical and physical characteristics of the catalog of identified ADF--S sources,
and a detailed description of the procedures for identifying counterparts in external catalogs was presented in \cite{malek10}. 

\subsection{Spectroscopic redshift information} 
\label{zsection}
The redshift information was originally only available for  48 galaxies among the 545 galaxies identified in \cite{malek10}; 
for an additional 5 galaxies, we found redshifts in NED after first publication.
Additionally, we used spectroscopic redshifts for ADF--S sources observed at 90$\mu$m, measured by \cite{sedgwick11}.
Combining data from \cite{malek10}, \cite{sedgwick11}, and new data from the NED and SIMBAD databases, gives us 173 galaxies with known spectroscopic redshifts.

Nearly all of the sources in these sample are nearby galaxies, with a
mean redshift of $0.089 \pm 0.014$, and a median value of 0.058. 
The range ($0 < z < 0.25$) corresponds to the range used by \citet{sedgwick11} to calculate the far-infrared luminosity function of local star-forming galaxies.
\begin{figure}[t]
	\resizebox{0.9\hsize}{!}{\includegraphics{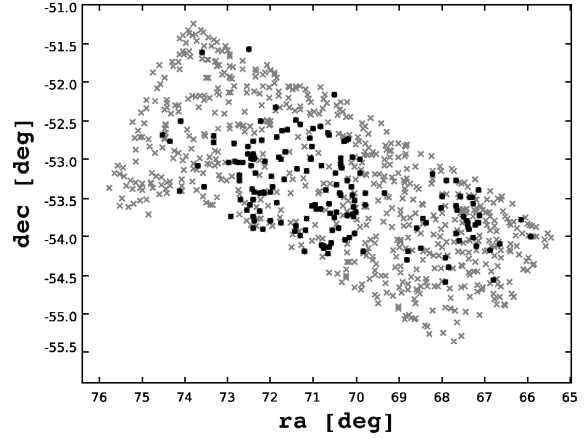}}
	\caption[]{Positions of 173 galaxies with spectroscopic redshifts information found in public databases and the Sedgwick et al. (2011) catalog. 
	These galaxies also belong to the sample of the 1 000 sources brighter than 0.0310 Jy from the ADF–S catalog that are analyzed in this paper. 
The first 1\,000 of  brightest sources from the ADF--S catalog in WIDE--S band are marked  with gray $\times$ symbols. 
Sources with spectroscopic redshifts are shown as full circles.}
	\label{SPECmapredshifts}
\end{figure}

The positions of all sources with known spectroscopic redshifts are shown in Figure~\ref{SPECmapredshifts}. 
Two separated areas rich in spectroscopic redshifts are clearly visible: (1) centered in the right part of Figure~\ref{SPECmapredshifts}, connected with lenticular-rich galaxy cluster Abell S0463 at z $\sim$ 0.039, and (2) located in the middle part of ADF--S field, observed by \cite{sedgwick11}.

\section{Spectral energy distribution fitting for sources with known spectroscopic redshift}
\label{sedsection}

\subsection{Sample selection} 
\label{selection}
To study physical parameters of ADF--S sources we selected from our sample 95 galaxies with known spectroscopic redshift and with the highest quality photometry available to fit SED models. 
The main criterion  was to have redshift information, and at least six measurements spanning the galaxy spectra. 

\begin{figure}[t]
	\resizebox{0.9\hsize}{!}{\includegraphics{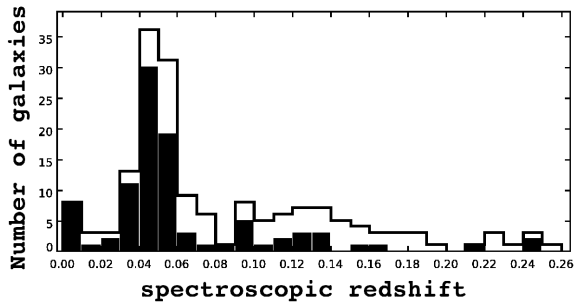}}
	\caption[]{The spectroscopic redshift distribution, in 0.01 bins.
	The open histogram corresponds to the distribution of 173 galaxies with known spectroscopic redshifts in the sample of the 1\,000 brightest ADF--S galaxies. 
	One object with redshift higher than 0.3 is not shown (quasar HE 0435$-$5304, located at z=1.232).		
	The solid histogram corresponds to the 95 galaxies used for SED fitting.
	}
	\label{SPEC}
\end{figure}

The redshift distribution for this selected sample is presented in Figure~\ref{SPEC}. 
The mean redshift  is  $0.062 \pm 0.045$, and the median is 0.046. 
{ It implies that our sample consists of nearby galaxies; their high IR flux is related to their intrinsic physical properties.}  

The distribution of the six measurements is tightly related to our sample selection. 
Each galaxy has at least one measurement in the FIR wavelengths from the AKARI FIS 90$\rm{\mu}$m detector and optical part of the spectra. 
{ Only five galaxies (5.29\%) have no measurements from the WISE survey (MIR bands); 
85.26\% galaxies from our sample are detected in all 2MASS bands (J, H, K$\rm{_s}$);  
half of our sample was also detected in the UV (FUV and NUV bands from GALEX survey);  
23 galaxies have at least one IR measurement in the IRAS database. 

For the whole sample the data cover the optical to FIR part of spectrum.
Missing FUV measurements for one half of the sample might result in a worse fitting for the spectra, but additional tests with and without GALEX data confirmed the homogeneous quality of fitted spectra in our sample. 
We concluded that the obtained results can be  used for analysis with the same significance for the whole sample. 
In most cases the lack of  UV measurement is probably caused by the selection effects: at the measured redshift, the expected UV flux is too low to be detected by GALEX. 
}

{ The percentage of available measurements in NIR, MIR, and FIR suggests that our sample consists of galaxies that are bright in IR and rich in dust. 
A high ratio of sources visible in UV implies ongoing star formation processes in these galaxies (i.e., the presence of numerous young stars emitting in UV). }
All available measurements from the ADF--S database were used {\bf for} the SED fitting.

\subsection{CIGALE}
The CIGALE \citep{noll09}, SED-fitting program, was used to estimate physical properties of galaxies. 
Very briefly, CIGALE  builds a stellar population model, then the UV-optical part of the spectra is reddened by an attenuation law, and the energy absorbed by dust is re-emitted in the IR \citep{noll09}. 
In that way, CIGALE computes all possible spectra, and calculates mean fluxes in the observed filter bands.
Bayesian-like statistical analysis (details can be found in \citealp{noll09}) permits the estimation of the best-fit model for each galaxy 
from $\chi^2$ minimization, and the $\chi^2$ value determines the quality of this SED fit. 

{The comparison between the model and the photometry for selected filters, i.e. the comparson of $\rm{f_{mod,i}}$ (per $\rm{M_{\odot}}$) and $\rm{f_{obs,i}}$ for $N$ filters, is carried out for each model by the minimization of} 

\begin{equation}
 \rm{\chi^2(M_{gal})=\varSigma_{i=1}^{N} \frac{ (M_{gal}f_{mod,i} - f_{obs,i} )^{2} } {\sigma_{obs,i}^{2} },  }
\end{equation}
{ where the galaxy mass ($\rm{M_{gal}} [M_{\odot}]$) is a free parameter, and $\rm{\sigma}$ describes a statistical photometric error \citep{noll09}. 
Uncertainties of photometric measurements are taken into account during fitting process.
In the next step CIGALE derives model-related probability from reduced best-fit  $\rm{\chi^2}$, and finally computes expected values and standard deviations for every parameter, based on the probability-weighted distribution of the model parameter values \citep{noll09,cigale_help}. 

The resultant value of $\rm{\chi^2_{reduced}}$ calculated for each model is then used to identify the model corresponding to the best-fit SED in the entire grid of models used in the computation. 
}

Thus CIGALE models the emission from a galaxy in the wavelength range from the  far-UV to the FIR. 
Models of emission from stars are given either by \citet{maraston05} or \citet{fioc97}.
For the calculation of initial mass function (IMF) CIGALE has two built-in algorithms: \citet{salpeter55} and \citet{kroupa01} IMFs.
The absorption and scattering of star light by dust, the so-called attenuation curve for galaxies adopted by CIGALE is based on the Calzetti law \citep{calzetti00} with some modifications. 
The slope of the dust attenuation is controlled by the factor $(\lambda/\lambda_{0})^{\delta}$, where $\lambda_{0}$ is the normalization wavelength, and $\delta$ the slope of  the modifying parameter \citep{noll09, buat11, boquien12}.
With regard to the base of dust attenuation, therefore $\delta$=0 corresponds to a starburst attenuation curve, $\delta>0$ to a shallower one, and $\delta<0$ to a steeper one, as in the Magellanic Coulds \citep{boquien12}.  
A second possible modification is to add a UV bump to the attenuation law, with its amplitude as a free parameter \citep{noll09}. 

Dust emission is calculated by a model proposed by \citet{dale02}. 
In this model, the IR part of SEDs is given by a power-law distribution
\begin{equation}
 dM_{d}(U) \propto U^{-\alpha_{SED}}dU,
\end{equation}
where $\rm{M_d(U)}$ is the mass of dust heated by a radiation field $\rm{U}$, and $\alpha_{\rm{SED}}$ is the heating intensity. 
Power-law representations for $\rm{dM_{d}(U)}$ are based on two extremes of heating: 
1) diffuse heating, where the intensity falls off  away from the heating source, and 2) a dense medium where the heating intensity is primarily attenuated by dust absorption \citep{dale01}.
The $\alpha_{\rm{SED}}$ parameters proposed by \citet{dale02} range from 0.0625 to 4, as the dust heating ranges from strong to quiescent \citep{dale05}.
Templates with $\alpha_{\rm{SED}}\leqslant1$ represent very active star-forming galaxies.
Normal galaxies have $1< \alpha_{\rm{SED}} <2.6$, where $\alpha_{\rm{SED}} \sim 2.5$ means quiescent \citep{dale01,dale02}.
Galaxies with $\alpha_{\rm{SED}}$ higher than 2.5 seem to have their FIR emission peak at even longer wavelengths than most quiescent galaxies ever studied \citep{dale05}.
In other words, \citet{dale02}  describes the progression of the far-infrared peak toward shorter wavelengths for increasing global heating intensities \citep{dale01}.  
 
The CIGALE code uses 64 templates for $\alpha_{\rm{SED}}$ values from 0.0625 to 4.0. 
After a number of tests, we decided to use eight of them in our analysis. 
The list of input parameters of the CIGALE code is shown in Table~\ref{cigale_parameters}. 
{ Redshift (either spectroscopic or photometric) is also used as an input parameter in SED fitting. }
We adopt a model of emission from stars given by \citet{maraston05}, with an IMF described by \citet{kroupa01}. 
The star formation history used in CIGALE is the combination of models for young and old populations. 
In our analysis we adopted an exponentially decreasing star formation rate for the old stellar population, starting 13 Gyr ago, and constant SFR (the so-called box model) for the young stars, starting 0.025 to 1 Gyr ago.  

\begin{table*}[t]
\begin{center}
\caption[]{List of the input parameters of the code CIGALE, based mostly on \cite{buat11}.}
\begin{tabular}{l | l | l}
\label{cigale_parameters}
Parameter &  Symbol & Values \\ \hline \hline
metalicities (solar metalicity) & Z & 0.2        \\                    
ages of old stellar population models in Gyr  & $t_{1}$ & 13\\
ages of young stellar population models in Gyr  &  $t_{2}$ & 0.025 0.05 0.1 0.3 1.0\\
fraction of young stellar population &  $f_{yS P}$ & 0.001 0.01 0.1 0.999 \\ \hline
slope correction of the Calzetti law &  $\delta$ & -0.3 -0.2 -0.1 0. 0.1 0.2\\
V-band attenuation for the young stellar population  & $A_{V,yS P}$ & 0.15 0.30 0.45 0.60 0.75 0.90 1.05 1.2 1.35 1.5 1.65 1.8 1.95 2.10 \\
Reduction of $\rm{A_{V}}$ basic for old SP models  & $f_{att}$ & 0.0 0.5 1.0  \\ \hline
IR power-law slope &  $\alpha_{\rm{SED}}$&  0.5 1.0 1.5 1.75 2.0 2.25 2.5 4.0  \\     
$\rm{L_{dust}}$ fraction of AGN model & $f_{AGN}$ & 0. 0.1 0.25 \\ \hline
\end{tabular} 
\end{center}
\end{table*}

\begin{figure*}[t]\centering
 	\resizebox{1\hsize}{!}{\includegraphics{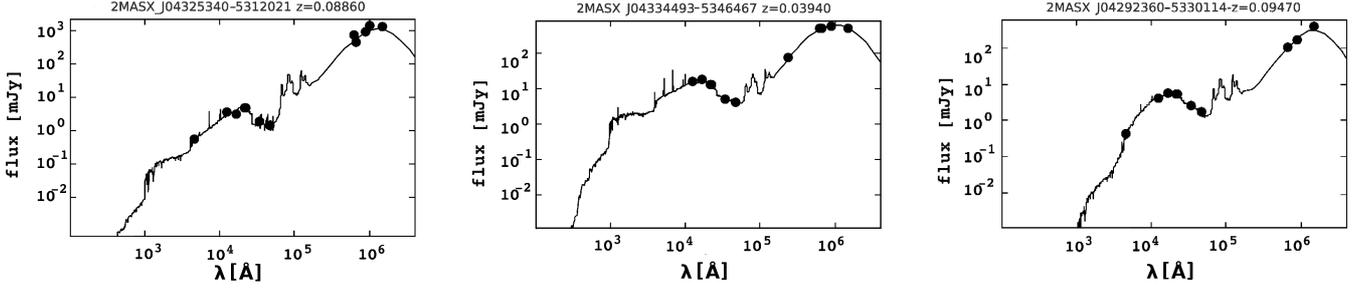}}
        \caption{Three examples of the best-fit models. 
Solid lines correspond to the best model obtained from CIGALE. 
Full black circles represent observed data used for SED fitting.}
    \label{SEDy}
\end{figure*}

The CIGALE code provides results for more than ten parameters based on SED model-fitting, using as a scaling parameter galaxy mass ($\rm{M_{\mathrm{gal}}}$). 
Examples of the best-fit models obtained from CIGALE are given in Figure~\ref{SEDy}. 

\subsection{Reliability test}
One possible test of reliability of the parameter estimation by CIGALE was described by \citet{giovannoli11}, and later used by \citet{buat11}, \citet{yuan11}, and  \citet{boquien12}.  
The strategy is based on building a mock cataloge.

To construct the mock cataloge, first it is necessary to compute  the best SED fit for each observed galaxy. 
We then take for further analysis galaxies with the best $\rm{\chi^2_{reduced}}$ value lower than four. 
This threshold value was chosen as a mean $\rm{\chi^2_`{min}}$ value + 3 $\sigma$, and our choice was confirmed by a visual inspection of all SEDs fitted by CIGALE to our spectroscopic sample. 
We added to their observed fluxes a random uncertainty drawn from Gaussian distribution with $\sigma$ = 0.1 (typically 10\% of flux value).
Thus we obtained a mock cataloge with the flux information for every photometric band used for our analysis.   
Then, the last step of checking estimated parameters is to run CIGALE code on the mock sample, using the same set of input parameters as in the first iteration, and compare the parameters of the artificial cataloge and analyzed observed sample.  

\begin{figure*}\centering
 	\resizebox{1\hsize}{!}{\includegraphics{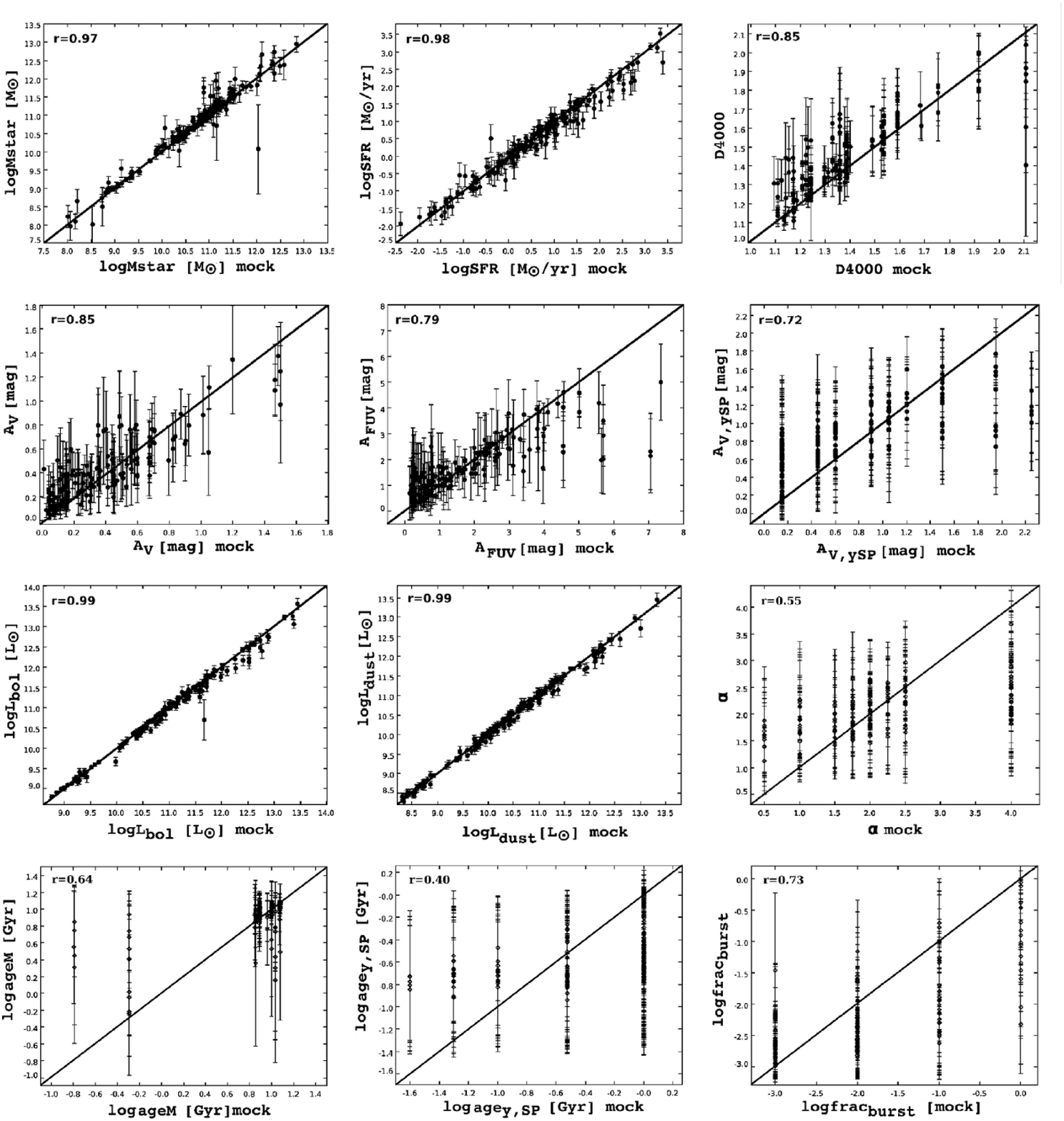}}
        \caption{Output parameters of Bayesian analysis of the  mock cataloge (x-axis) and the results of SEDs fitting for ADF--S galaxies (y-axis). 
        { Uncertainties calculated by CIGALE are shown on the vertical axes.}        }
      \label{mock_cataloge}
\end{figure*}

Figure~\ref{mock_cataloge} compares the output parameters of the mock cataloge versus values estimated by the code. 
The comparison between the results from the mock and real cataloges shows that CIGALE gives a very good estimation of parameters concerning 
\begin{enumerate}
 \item star formation history, based on the \citet{maraston05} model and IMF from \citet{kroupa01}: 
\begin{itemize}
 \item stellar mass (M$_{\mathrm{star}}$);
 \item star formation rate (SFR);
 \item the age-sensitivity D4000 index defined as the ratio between the average flux density in the 400--410nm range and that in the 385--395 nm range \citep{boquien12}.
 \end{itemize}
 \item dust attenuation, based on modified \citet{calzetti00} law: effective obscuration factors in magnitudes at 1500 $\pm$ 100 \AA{} ($\mathrm{A_{FUV}}$) and 5500 $\pm$ 100 \AA{} ($\mathrm{A_{V}}$), and  dust attenuation in the V band for the young stellar population ($\mathrm{A_{v,ySP}}$),
 \item dust emission given by \citet{dale02}: bolometric and dust  luminosity ($\mathrm{L_{bol}}$, $\mathrm{L_{dust}}$). 
\end{enumerate}
All the parameters listed abovewere estimated with values of the linear Pearson moment correlation coefficient, (r), higher than 0.8.
 
The accuracy of the stellar mass fraction due to the young stellar population ($\mathrm{frac}_\mathrm{burst}$), mass-weighted age of the stellar population ($\mathrm{age}_\mathrm{M}$), and  heating intensity from the \citet{dale02} models $\alpha_\mathrm{SED}$ are estimated with slightly lower efficiency (correlation coefficient equal to 0.73, 0.64, and 0.55, respectively). 
The age of young stellar population ($age_\mathrm{ySP}$) is poorly constrained  with $r$=0.40.
The Pearson product-moment correlation coefficient ($r$) is shown in the plots (Fig.~\ref{mock_cataloge}). 

Based on the accuracy of the CIGALE code output parameters, estimated from the coefficient value $r$, we decided to study only $\mathrm{M_{star}}$, SFR, D4000, $\mathrm{A_{V}}$, $\mathrm{A_{FUV}}$, $\mathrm{A_{V, ySP}}$, $\mathrm{L_{bol}}$, $\mathrm{L_{dust}}$, and $\mathrm{frac_{burst}}$, with a brief 
consideration of $\mathrm{age_{M}}$ and starburst activity related to the $\alpha_\mathrm{SED}$ value.

\section{Results}
\label{resultssectionSPECT}
	\begin{figure}[t]
		\resizebox{0.9\hsize}{!}{\includegraphics{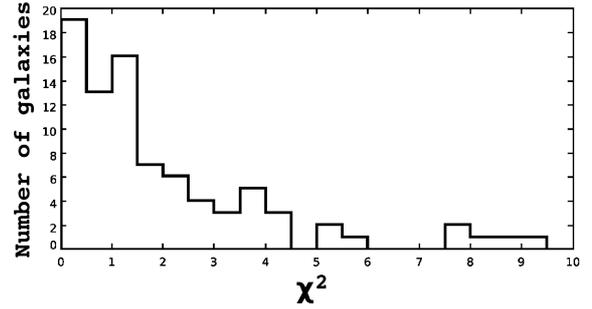}}
		\caption[]{The distribution of the $\rm{\chi^2_{reduced}}$ values obtained from the Bayesian analysis of 95 galaxies with known spectroscopic redshifts that were used for SED fitting.
Eleven galaxies with $\rm{\chi^2_{reduced}}$ larger than 10 are not shown in this plot. 
{ For 32 galaxies $\rm{\chi^2_{reduced}}$ is lower than 1, with mean and a median value equal to 0.42.}}
		\label{chi2_histo}
	\end{figure}

\begin{figure}[t]\centering
	\resizebox{1.0\hsize}{!}{\includegraphics{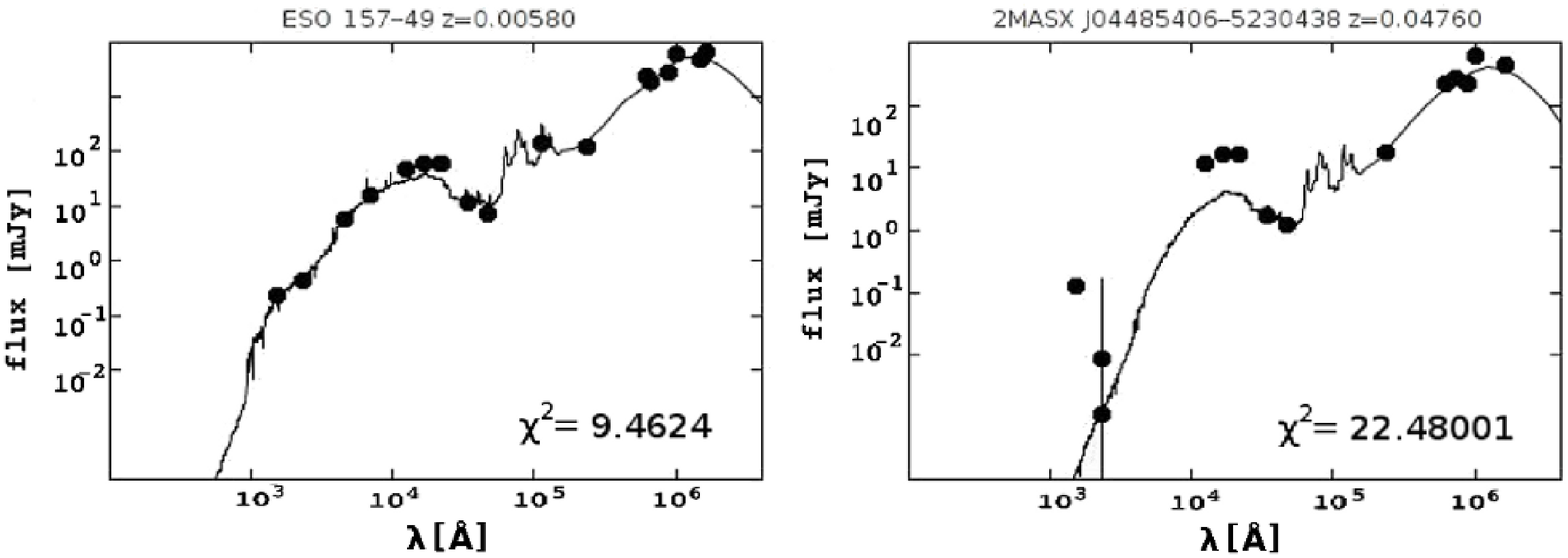}}
         \caption{{Two examples of the poorly fitted spectra, with $\rm{\chi^2_{reduced}}$ higher than four.
         In the first case (left panel) the $\rm{\chi^2_{reduced}}$ value is still reasonable, and  the physical parameters of the best model probably are close to real ones. 
         In the case of a second example (right panel), the model fits mainly the dusty part of the spectrum. }}
      \label{badSEDs}
\end{figure}

The distribution of 95 $\rm{\chi^2_{reduced}}$ values obtained from the Bayesian analysis is plotted in Figure~\ref{chi2_histo}.
We restrict the analysis to the SEDs with a best-fit value of $\rm{\chi^2_{reduced}}$ lower than four, which is 73 out of the 95 galaxies used for SED fitting. 
{ The remaining 22 galaxies show a much worse conformity to the model. 
As shown in Fig.~\ref{badSEDs}, it might be caused by a huge uncertainty of measurements, or even inhomogeneities in the collected data points. 
The majority of these galaxies (17 sources) have angular sizes larger (several or even tens of times) than the aperture size used for the AKARI FIS for 90$\mu$m flux measurement, which implies that their FIR measurements are significantly underestimated, and - to a different degree - it also applies  to measurements at other wavelengths. 
It increases our confidence that the $\rm{\chi^2_{reduced}}$ threshold method works well, and all suspicious sources were rejected during SED fitting process. 
All these galaxies are very near, but they do not represent any special class from the point of view of the physical parameters, so their rejection from the sample should not bias the final results. }

{ The mean angular distance of the nearest counterpart from the ADF-S source for the final sample of 73 sources with spectroscopic redshifts that were used for the analysis is equal to 6.70'', while the corresponding median value is equal to 6.27''. 
The maximum angular distance between the ADF--S source and its optical counterpart is equal to 17.82'' (corresponding to the 40\% of the aperture radius for 90$\mu$m AKARI band). 
This small positional scatter between AKARI sources and their counterparts (see Malek et al., 2010), as well as the careful visual inspection of all sources, allows us to expect only a small fraction of possible chance coincidences in our sample. 
}

    \begin{figure*}[t]\centering
	\resizebox{1\hsize}{!}{\includegraphics{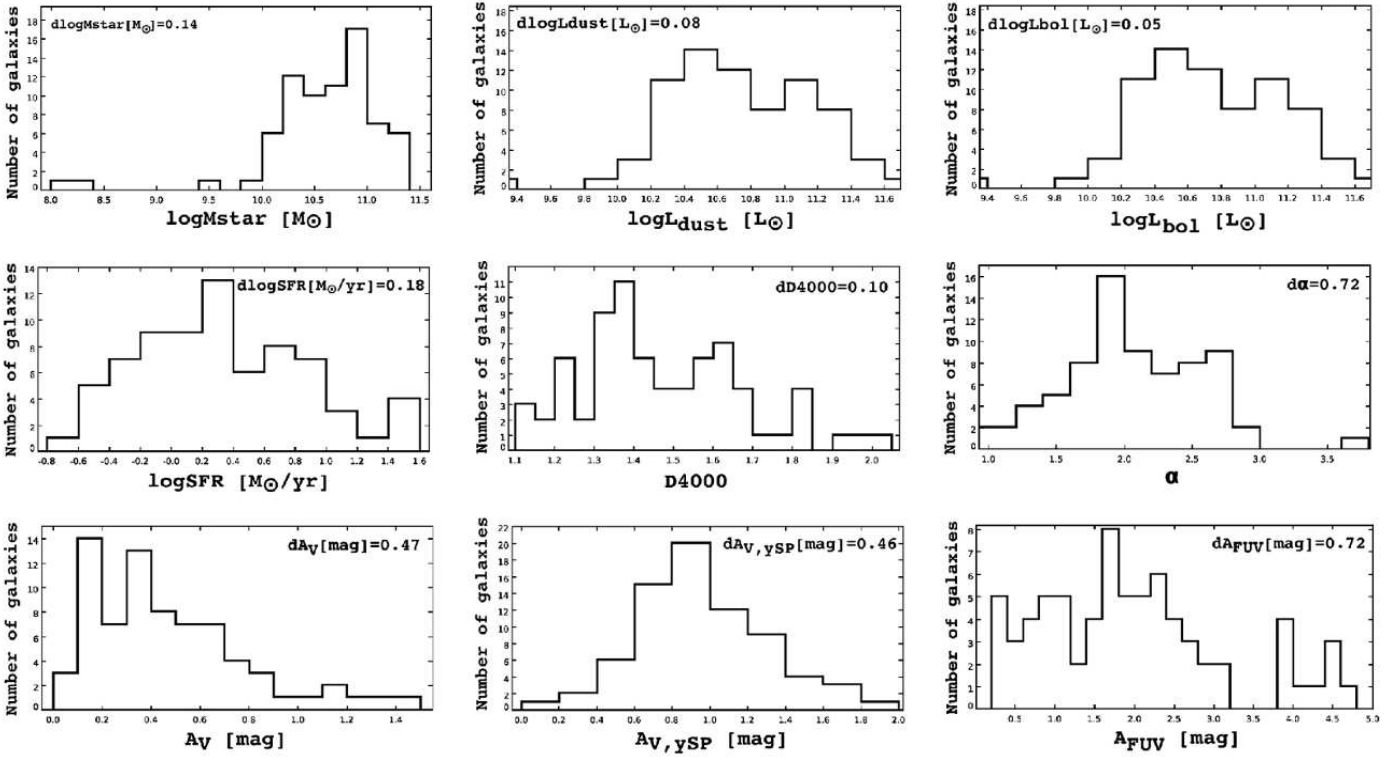}}
         \caption{Distribution of the Bayesian estimates of the output parameters discussed in this work:  $\mathrm{M_{star}\mbox{ }[M_{\odot}]}$,  $\mathrm{L_{dust}}$ and $\mathrm{L_{bol}\mbox{ }[L_{\odot}]}$, $\mathrm{logSFR\mbox{ }[M_{\odot} yr^{-1}]}$, $\alpha_\mathrm{SED}$, $\mathrm{A_{V}}$ [mag],  $\mathrm{A_{V,ySP}}$ [mag], $\mathrm{A_{FUV}}$ [mag], and D4000.
         { Mean standard deviation for each physical parameter is shown as \texttt{dP}, where \texttt{P} is the name of the parameter. }
Only galaxies with $\chi^{2}$ less than four are presented.}
      \label{FigGam_all}
    \end{figure*}

The distribution of the main parameters estimated with the Bayesian analysis for 73 galaxies in our sample is plotted in Figure~\ref{FigGam_all}.
We found a median stellar mass  of our sample equal to  $\mathrm{M_{star} = 4.83 \cdot10 ^ {10}\mbox{ }[M_\odot]}$.
The analyzed sample is  rather bright, $\mathrm{L_{bol} = 5.19 \cdot 10 ^ {10} \mbox{ }[L_\odot]}$, and also has high dust luminosities,  $\mathrm{L_{dust} = 1.47 \cdot 10 ^ {10} [L_\odot]}$,  
with minimum and maximum value of $\mathrm{L_{dust}}$ equal to 8.73 and 11.51 $\mathrm{[L_\odot]}$, respectively. 
This means that part of our sample consists of luminous infrared galaxies (LIRGs).
Calculated values of star formation rate vary from 0.22 to 37.29 $\mathrm{[M_\odot yr^{-1}]}$, with the median value equal to 1.9 $\mathrm{[M_\odot yr^{-1}]}$. 
The value of the heating intensity $\alpha_\mathrm{SED}$ implies that the vast majority of analyzed galaxies (87.91\%) belong to the population
of normal, but rather active, star-forming galaxies, with a median value of $\alpha_\mathrm{SED}$ equal to 2.11. 
The median value of the $\mathrm{A_{V}}$ parameter, describing dust effective attenuation for stellar population at wavelength equal 5500 $\AA{}$ is 0.39 [mag], and the median value for attenuation in FUV (at 1500 \AA{}, $A_\mathrm{FUV}$) is 1.773 [mag].
The parameter $\mathrm{A_{V,ySP}}$, which describes V-band attenuation for young stellar population models, spans almost the  entire range of input parameters from 0.1 to 1.8, with the median value 0.95 [mag].
Roughly these extinction parameters are expected, based on the observation that the typical dust luminosities measured in the infrared
are about one third of the total bolometric luminosities.

{
\subsection{Uncertainties of the physical parameters}

The CIGALE code derives properties of a galaxy and the associated uncertainties by analyzing the probability distribution function for each parameter. 
The code generates the output parameter file in which deviations from measured 
fluxes in all fluxes, errors used in the fitting process, weighted mean values, and standard deviations for all physical parameters are listed 
\citep{cigale_help}.
To compute uncertainties for the parameters CIGALE uses probability distribution functions (PDFs) described in detail in \cite{noll09}.   

Uncertainties obtained for the physical parameters by CIGALE are shown in Figure~\ref{mock_cataloge} on the vertical axes.
Additionally, the mean uncertainties of the parameters used in our analysis are shown in Figure~\ref{FigGam_all} as a \texttt{dP}, where \texttt{P} is the name of parameter. 

}
\subsection{Mass - metallicity relation}

During SED fitting we assumed a constant value of metallicity equal to 0.02 (similar to the solar metallicity, $Z_{\odot}$). 
However, based on the mass versus gas-phase metallicity relationship given by \cite{tremonti04}, we may estimate this value for every single galaxy, using  the formula 

\begin{multline}
\mathrm{12+log(O/H)=-1.492+1.847(log\mathrm{M_{star}})-}\\\mathrm{+0.08026(log\mathrm{M_{star}})^2}
\end{multline}
 
To obtain this formula, \cite{tremonti04} examined the sample of $\sim$ 53\,000 SDSS star-forming galaxies, with redshift cut $0.005<z<0.25$, finding a tight correlation between $\mathrm{M_{star}}$ and metallicity in the stellar mass range  $8.5<\mathrm{logM_{star}}<11.5\mbox{ }[M_\mathrm{\odot}]$. 
Our sample (except for two galaxies with $\mathrm{M_{star}}$ equal to 8.08 and 8.29, respectively), fulfills both conditions, i.e., in redshift and stellar mass range. 

The mean value of 12+log(O/H) from this relation for our sample equals to $9.06 \pm 0.06$, (with median 9.08), and indicates that our galaxies are high-metallicity galaxies, defined as $\rm{12+log(O/H)>8.35}$ \citep{calzetti07}.  
For comparison, the solar metallicity is lower and equals to $\rm{12 + log(O/H) = 8.69 \pm 0.05}$ dex \citep{allende01}.
We conclude that this high-metallicity is related to high luminosity in IR, and also with high dust content of analyzed sources. 
Similar conclusions may be found in \cite{yuan11}, based on the GALEX-SDSS-2MASS-AKARI (IRC/Far-Infrared Surveyor) sample of nearby galaxies (z$<$0.1), selected in IR. 

\subsection{Is our sample actively star forming?}

Our sample of 73 galaxies is characterized by a rather high SFR parameter, shown in Figure~\ref{FigGam_all}.
Fitted values of SFR range between 0.22 and 37 with a well-defined maximum and a median value of 1.9 $\rm{[M_{\odot}yr^{-1}]}$. 

\begin{figure*}[t]
  \begin{minipage}[t]{0.49\textwidth}
	  \resizebox{0.85\hsize}{!}{\includegraphics{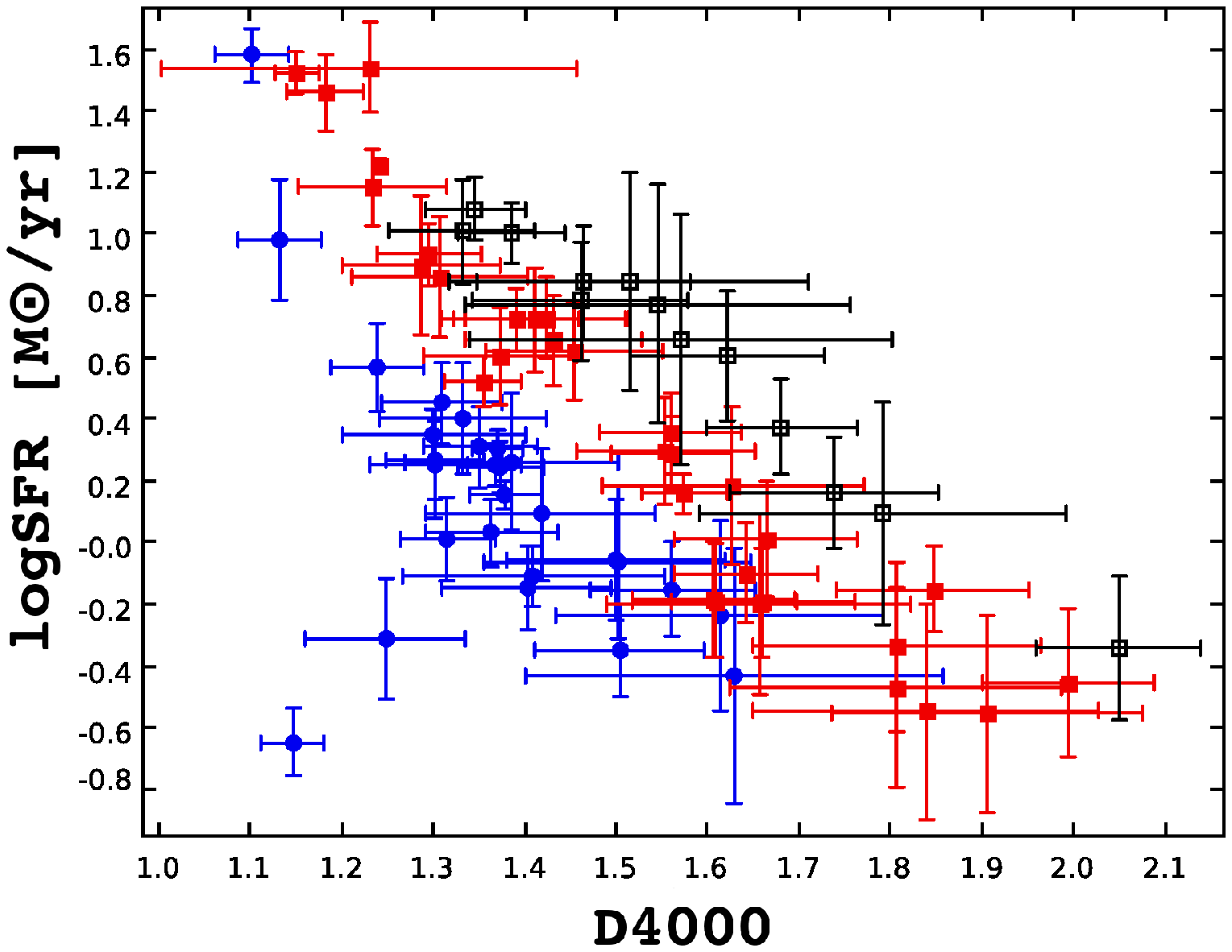}}
	  \caption[]{The relation between D4000 and SFR. 
  Open squares represent galaxies with $\mathrm{11 < logM_{star} <11.5}$.
  Galaxies with $\mathrm{log\mathrm{M_{star}}}$ in the range from 10.5 to 11  are marked as solid squares.
  Filled circles represent galaxies with $\mathrm{logM_{star}}$ less than 10.5.}
	  \label{D4000_vs_SFR}
  \end{minipage}\hfill 
  \begin{minipage}[t]{0.49\textwidth} 
	\resizebox{0.9\hsize}{!}{\includegraphics{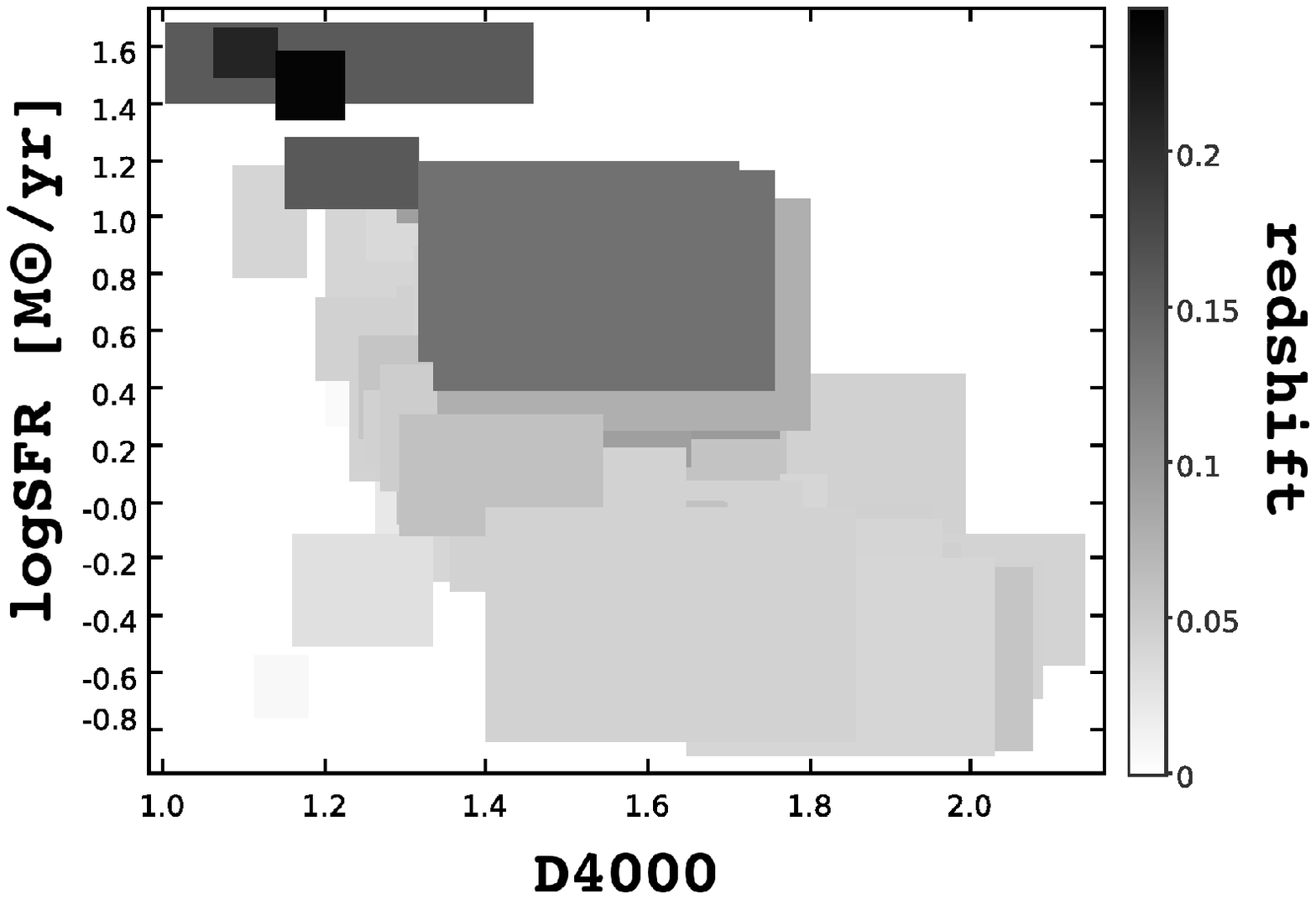}}
	\caption[]{The distribution of the Bayesian estimates of the logSFR parameter as a function of D4000 and redshift.
	The uncertainty ranges are shown as rectangles.}
	\label{z_vs_SFR}
	\end{minipage} 
\end{figure*}

The D4000 $\AA{}$ break (D4000, depth of Balmer break), defined as the ratio of the flux in the red continuum to that in the blue continuum \citep{balogh99}, may provide information about the age of stellar population. 
The break in spectra is caused by the accumulation of a large number of spectral lines in a narrow wavelength region, especially from metals.
Thus the D4000 break will be large for old galaxies, rich in metals, and smaller for young stellar populations \citep{kauffmann03}.
\cite{kauffmann03} found that the distribution of stellar mass shows a clear division between galaxies dominated by old  and more recent star formation, discriminated by D4000: 
D4000 $\sim$ 1.3 describes young galaxies, while old, elliptical galaxies are concentrated around a typical D4000 value of 1.85.  
Most objects in our sample (58.9\%) have low values of D4000 ($<1.5$).
Galaxies with D4000 lower than 1.3 comprise 17.8\%.
This suggests that our sample consists of dust-rich star-forming spiral {or lenticular} galaxies. 
We have checked the morphology distribution for all 73 galaxies in our sample; for half of them the classification is given in the NED database; we have found that 50\% of our sample belongs to the spiral, and 49\% to lenticular galaxies, respectively.
The result is consistent with \cite{malek10}.
We found that D4000 is inversely correlated with SFR (see Figure~\ref{D4000_vs_SFR}). 
A linear correlation with stellar mass is also noticable: the more the massive galaxy is, the higher its SFR is at the same D4000 level.

A similar correlation between SFR and redshift is shown in Figure~\ref{z_vs_SFR}. 
More distant galaxies in our sample are characterized by higher SFR and lower value of D4000, because in a flux-limited   sample, they are the more luminous objects.

\begin{figure}
		\resizebox{0.9\hsize}{!}{\includegraphics{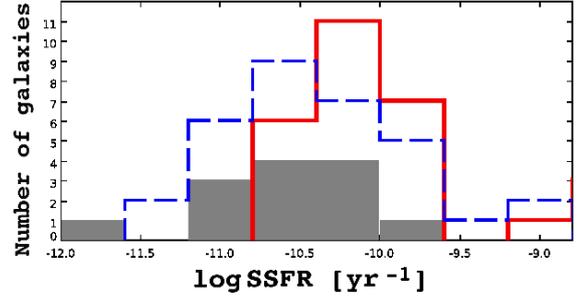}}
		\caption[]{The distribution of the specific star formation rate (SSFR).
The open red histogram (solid line) corresponds to galaxies with $\mathrm{logM_{star}<10.5}$, 
the open blue histogram (dashed line) to galaxies with $\mathrm{logM_{star}}$ in the range from 10.5 to 11, and  filled gray histogram to galaxies with $\mathrm{11<log\mathrm{M_{star}}<11.5}$.}
		\label{ssfr_histo}
	\end{figure}

We calculated the ratio of the current SFR to the total stellar mass,   called the specific star formation rate (SSFR=SFR/$\mathrm{M_{star} [yr^{-1}]}$).
The distribution of SSFR is shown in Figure~\ref{ssfr_histo}. 
The volume-weighted distribution of the SSFR was used to obtain the average trend of our sample, to compare with  other previously published \citep{buat07,brinchmann04} results. 
The distribution of weighted SFR as a function of $\mathrm{M_{star}}$ was calculated into 0.4 $\mathrm{logM_{\odot}}$ bins using the formulae 

\begin{equation}
 \textlangle X \textrangle = \frac{\sum_{i} \omega_{i} X_{i}}{\sum_{i} \omega_{i}},
\label{sredniawazona}
\end{equation}

\begin{equation}
 \sigma^{2}=\frac{\sum_{i} \omega_{i}[X_{i}- \textlangle X \textrangle]^2}{\sum_{i} \omega_{i}},
\label{sigmasredniawazona}
\end{equation} where $< X >$ is the average value of SSFR, $\omega_{i}$ is a weight, and  $X_i$ is equal to SSFR for each bin, respectively; 
$\omega_{i}$ is defined as ${1}/{V_{\mathrm{max}i}}$ for each galaxy, where $V_{\mathrm{max}i}$ is a volume defined as 
\begin{equation}
 V_{\mathrm{max}i}=\frac{4}{3}\pi(\frac{L_{\mathrm{bol}}}{4\pi F_\mathrm{WIDE--S}})^{3/2}.
\end{equation}

\begin{figure}[t]
	\resizebox{0.9\hsize}{!}{\includegraphics{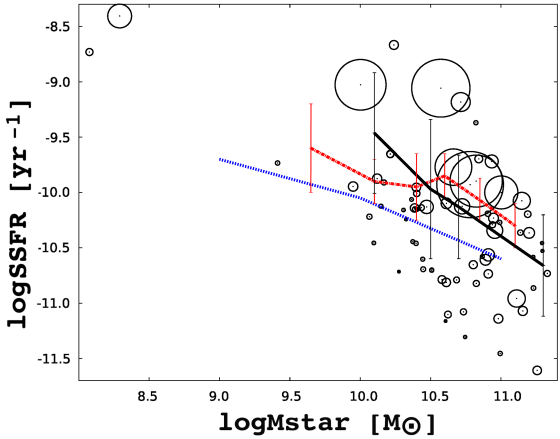}}
	\caption[]{Relation between the specific star formation rate (SSFR) and stellar mass {(the so-called star formation main sequence)}.
	The  solid black line represents the volume-weighted average values of SSFR as a function of $\mathrm{M_{star}}$ for our sample of 73 galaxies. 
	Error bars correspond to the volume-weighted standard deviation. 
	The stellar mass intervals are $\mathrm{logM_{star} \in [9.9-10.3], [10.3-10.7], [10.7-11.1],\mbox{ } and [11.1-11.5]}$. 
	The radius of the circle is related to the flux detected at the WIDE--S 90$\mu$m band. 
	The distribution of fluxes shows that the lower boundary of the stellar mass versus SSFR relation corresponds to the AKARI detection limit. 
	The average SSFR versus $\mathrm{logM_{star}}$ found by \citet{brinchmann04} is plotted as a dashed line. 
	The average SSFR for the FIR-selected sample obtained by \citet{buat07} and 1$\sigma$ errors are plotted as a dashed-dotted line.
}
	\label{Mstar_vs_SSFR_BB}
\end{figure}

We conclude, that the majority of our sample galaxies are actively forming stars. 
We also find that SSFR decreases with increasing stellar mass, and this trend is also visible even without applying the volume - weighted average (see Figure~\ref{Mstar_vs_SSFR_BB}). 
Our results are consistent with previous results  \citep[e.g.,][]{cowie96,brinchmann04,buat07,iglesias07}.
  
\citet{brinchmann04} performed  an analysis of the physical properties of star-forming galaxies using a sample of $\sim10^5$ galaxies with measurable star formation in the SDSS in the redshift range from 0.005 to 0.220.
\citet{buat07} presented results analogous to \citet{brinchmann04} for nearby galaxies selected in the far infrared and far ultraviolet: they found that SSFR decreases with increasing stellar mass {(star formation main sequence)} both for FIR and UV-selected sample. 
A similar trend appears in all the presented samples (see Figure~\ref{Mstar_vs_SSFR_BB}), however, in our sample, the SSFR is higher than the results obtained by \citet{brinchmann04} and \citet{buat11} for galaxies with $\mathrm{logM_{\odot}}$ $\sim$ 10 [$\mathrm{M_{\odot}}$].  
A slight disagreement with the previous estimations \citep{buat07,brinchmann04} may be caused by our small data set in the range of stellar mass below logM$_\mathrm{star}=10$. 
The radius of circles in Figure~\ref{Mstar_vs_SSFR_BB} represents the flux measured by AKARI at the WIDE--S 90$\mu$m band. 
Sources with the lowest fluxes have a tendency to occupy an area near the lower boundary on this plot.

The star formation main sequence result for our sample is higher than the  \cite{brinchmann04} result and partialy higher than the \cite{buat07} result, and this implies that the ADF–S sample has a relatively high SFR.

Simple checks in the NED database did not reveal any morphological nor environmental peculiarities for a group of galaxies characterized by low SSFR and high stellar mass seen in the the bottom-right corner in Figure~\ref{Mstar_vs_SSFR_BB}.  
 
The distribution of the heating intensity $\alpha_\mathrm{SED}$ shown in Figure~\ref{FigGam_all} confirms the conclusion that most of our sample consists of  normal, nearby galaxies, active in star formation.
Only 7.5\% of our galaxies have $\alpha_\mathrm{SED}$  higher than 2.6, which is the limit for normal galaxies given by \citet{dale02}.
The mean value of $\alpha_\mathrm{SED}$ for 92.5\% of galaxies with $\alpha_\mathrm{SED}<2.6$ is equal to  $1.9 \pm 0.7$, with a median value of 1.9, which confirms that our sample is not extremely active, but  still very bright in the FIR. 
Taking into account that $\alpha_\mathrm{SED}$ is not very well estimated by CIGALE, with a correlation coefficient equal to 0.55, it is difficult to draw explicit conclusions about starburst activity in our sample. 
However, the distribution of the D4000 parameter supports our view of substantial recent star formation.
 \cite{white12} also matched ADF--S sources with ACTA, a deep radio survey. 
The radio-luminosity plots they made suggest that the majority of radio sources observed by ACTA and having counterparts 
in the 90$\mu$m AKARI band  are luminous, star-forming galaxies \citep{white12}.

\section{Photometric redshifts}
\label{photo_subsection}

As explained in Sect.~\ref{zsection}, spectroscopic redshifts ($z_\mathrm{spec}$) are now available for 173 galaxies out of the 545 galaxies identified by \cite{malek10}. 
In our sample, 416 sources are identified in public catalogs as galaxies, with additional photometric information but no redshift.
To enlarge our sample, we estimated photometric redshifts ($z_\mathrm{phot}$) of the entire sample.
We used two codes: \textit{Photometric Analysis for Redshift Estimate} (Le PHARE, \citealp{arnouts99, ilbert06}), and \textit{Code Investigating GALaxy Emission} (CIGALE, \citealp{noll09}).

We performed a test on a  sample of 95 galaxies with known $z_\mathrm{spec}$ and at least six photometric measurements in the whole spectral range.

Le~PHARE estimates photometric redshifts based on a $\chi^2$ fitting method between the theoretical and observed photometric catalog. 
For our work we tested all libraries included in the Le~PHARE package. 
Based on our preliminary results we chose Rieke LIR templates \citep{rieke09} constructed for eleven luminous and ultraluminous purely star forming galaxies. 
This library is a part of the Le~PHARE code, and can be used to estimated photometric redshifts. 
Le~PHARE with the \citet{rieke09} galaxy template gave results for 88\% of galaxies in our sample; for 11 galaxies it was not possible to  estimate photometric redshift (a Le PHARE response equal to  $-99$ or gives redshift 0). 
Based on equations presented in \citet{ilbert06}, we found three galaxies with  
\begin{equation}
\Delta{z}=|z_\mathrm{spectro}-z_\mathrm{LePhare}|/(1+z_\mathrm{spec})>0.15, 
\label{CE}
\end{equation}
so-called catastrophic errors or catastrophic failures. 

{ \cite{ilbert06} used equation~\ref{CE} for the whole sample of 3\,241 galaxies from the VIMOS VLS Deep Survey in the spectroscopic redshift range 0-5. Later, \cite{ilbert13} obtained the same value of 0.15 for the sample of 9\,389 zCOSMOS bright galaxies of $\rm{K_s}<24$. 
The latter sample has a median spectroscopic redshift equal to 0.5 which shows that this criterion can  also  be appropriate for galaxies at lower redshifts.
}

\begin{figure}[t]
    \resizebox{0.99\hsize}{!}{\includegraphics{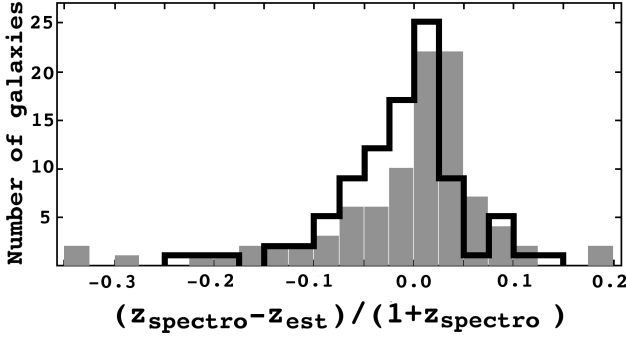}}
    \caption[]{Histogram of the differences between photometric redshifts obtained by Le PHARE ($z_\mathrm{LePhare}$, solid  black line) 
    or CIGALE ($z_\mathrm{CIGALE}$, gray histogram) and  spectroscopic redshifts for 95 ADF--S galaxies in 0.05 redshift bins.
    Eleven galaxies without successfully measured photometric redshifts ($z_\mathrm{LePhare}$)  are not shown in this histogram.
    }
   \label{zphoto_zspectra}
\end{figure}

Altogether, 14 galaxies (15\%) in our sample failed to have a successfully measured redshift. 
For 61\% of our galaxies the accuracy was better than $0.055(1+z_\mathrm{spectro})$, which was a condition for successful measurement adopted for the fainter part of a sample in \citet{ilbert06}. 
The distribution of $\Delta z$ is shown in Figure~\ref{zphoto_zspectra}.

\begin{figure}
    \begin{center}\resizebox{0.9\hsize}{!}{\includegraphics{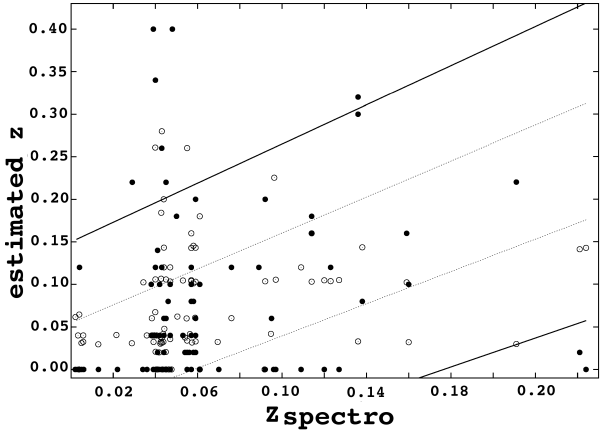}}\end{center}
    \caption[]{Photometric versus spectroscopic redshifts for a sample of 95 galaxies. 
    Open circles correspond to redshifts estimated using Le PHARE ($z_\mathrm{LePhare}$). 
    Black circles represent galaxies with $z_\mathrm{CIGALE}$ redshifts. 
    The region of catastrophic errors $\eta$, defined as $|z_\mathrm{spectro}-z_\mathrm{LePhare}|/(1+z_\mathrm{spectro})>0.15$ \citep{ilbert06}, is marked by a solid black line. 
    Dashed-dotted lines correspond to the $z_\mathrm{photo}=z_\mathrm{spectro}\pm0.055\cdot(1+z_\mathrm{spectro})$. 
    Eight galaxies without successfully measured photometric redshift ($z_\mathrm{LePhare}$)  are not included in this plot.
  }
  \label{zphoto_zspectra_lines}
\end{figure}	

A possible way to improve the estimation of photometric redshifts through including FIR data is provided by CIGALE. 
Originally, CIGALE was not developed as a tool for estimation of $z_\mathrm{photo}$, but since it uses a large number of models covering  the whole spectrum including IR, it may be expected to provide better $z_\mathrm{photo}$ than the software using mainly optical to NIR data, especially in the case of IR-bright galaxies.
The final redshifts assigned by CIGALE are based on the reduced best-fit  $\chi^2$ value of the fit.

In the same sample of the 95 galaxies which were also given to Le~PHARE, CIGALE found redshifts for all of them. 
Only seven of these turned out to have catastrophic errors, and even these were similar to them in the case of photometric redshifts given by Le PHARE (mean value of $|z_\mathrm{spectro}-z_\mathrm{LePhare}|/(1+z_\mathrm{spectro})$ for the catastrophic errors if equal to 0.22 $\pm$ 0.07 for CIGALE and 0.27 $\pm$ 0.02 for the Le~PHARE). 
In Figure~\ref{zphoto_zspectra_lines} all the estimated redshifts ($z_\mathrm{LePhare}$ and $z_\mathrm{CIGALE}$) are shown.

We found that less than 10\% of $z_\mathrm{CIGALE}$ may suffer from catastrophic errors, while for the same sample Le~PHARE was not able to determine redshifts for 12\% of sources and in an additional 3\%, the estimated values had catastrophic errors.
Results plotted in Figure~\ref{zphoto_zspectra_lines} show that catastrophic differences between estimated and spectroscopic redshifts 
are larger for  $z_\mathrm{LePhare}$ than  $z_\mathrm{CIGALE}$.
Consequently, we decided to use  $z_\mathrm{CIGALE}$ for sources without known spectroscopic redshift for the subsequent analysis.

{Following \cite{ilbert06}, we calculated the redshift accuracy $\rm{\sigma_{\Delta z}/(1+z_{spect})}$ for $z_\mathrm{CIGALE}$. 
The precision is equal to 0.056, much lower than those obtained by \cite{ilbert06}, but this should be attributed to much poorer statistic sof our sample (95 galaxies) when compared to \cite{ilbert06, ilbert13} (3\,241 and 9\,389 galaxies, respectively).
As shown in Figure~\ref{zphoto_zspectra_lines}, the highest uncertainties appear for the spectroscopic redshift range between 0.03 and 0.05, and we take this into account in our analysis. }

\begin{figure}[t]
	\resizebox{1\hsize}{!}{\includegraphics{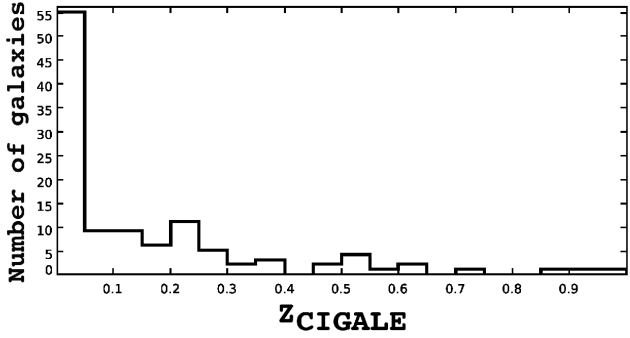}}
	\caption[]{The distribution of photometric ($z_\mathrm{CIGALE}$) redshift of 113 galaxies with a $\rm{\chi^2_{reduced}}$ value lower than 4 in 0.05 bins.
}
\label{histo_zcigale}
\end{figure}

We used CIGALE to calculate photometric redshifts with additional IR measurements for the remaining  galaxies without any a priori information about the distance. 
This procedure was possible for 127 galaxies with at least six photometric measurements. 
The distribution of $z_\mathrm{CIGALE}$ is shown in Figure~\ref{histo_zcigale}.
The mean value of $z_\mathrm{CIGALE}$ is $0.241 \pm 0.437$.  
The median value is equal to 0.097, but a high-redshift tail is clearly visible.  

\begin{figure}[t]
	\centering
	\resizebox{1\hsize}{!}{\includegraphics{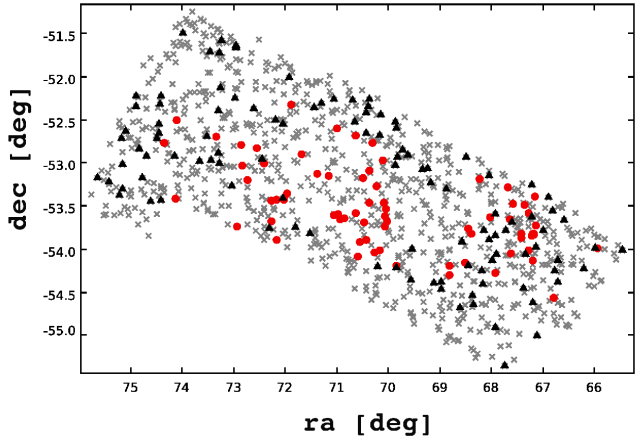}}
	\caption[]{The map of the first 1000 ADF--S sources brighter than 0.0301~Jy in the WIDE--S (90 $\mu$m) AKARI band. 
	Sources without information about redshift, sources for which SEDs fitting failed fitting ($\rm{\chi^2_{min}>4}$), and sources not identified are marked with a gray x symbol.  
	Galaxies with known spectroscopic redshifts are indicated by full red circles; sources with estimated $\mathrm{z_{CIGALE}}$ are filled black triangles.}
	\label{map_all_sources}
\end{figure}

In our work we decided to use all galaxies with known redshifts with $\rm{\chi^2_{reduced}}$ value of SED fitting lower than four. 
All galaxies with estimated photometric redshift are listed in Table~\ref{allzcigale} (page~\pageref{tableallzcigale} and following). 
The names of counterparts, the positions of the ADF-S sources, and the corresponding nearest counterparts, as well as the photometric measurements used for redshift estimation, are given [in Table] A.1, available at the CDS. abase of CDS. 
The map of positions of these sources on the sky is shown in Figure~\ref{map_all_sources}. 
{ The mean distribution of the angular deviations of the nearest counterparts from the photometric redshift sources used for the analysis is equal to 7.78'', with a median value 6.54''. 
The maximum distance between an ADF--S source and its optical counterpart is equal to 30.06''. 
For only 4\% of our photometric sample  (five sources), the distance between the  optical counterpart and its ADF-S coordinates is higher than 20''. 
Only ten sources in this sample (9\%) have angular sizes larger than the aperture radius of the AKARI 90 $\mu$m band, but the difference is small ($<$ 5''). As a result, the flux measured in the aperture may be underestimated, but the difference remains within the assumed flux uncertainties, and the final fits are not affected significantly by this difference. 
We also checked that in case of all the other measurements at different wavelengths gathered from different surveys, aperture sizes are usually close to the physical sizes of our sources.
We have checked the deviation of model filter fluxes in the infrared from the measured ones, and the $\rm{\chi^2_{reduced}}$ of SED fitting for all of them. 
The deviations for the WIDE--S band is close to zero ($\sim$0.01 mag), and $\rm{\chi^2_{reduced}}$ lower than theestablished treshold (four).}

\begin{figure}[t]
	\centering
	\resizebox{1\hsize}{!}{\includegraphics{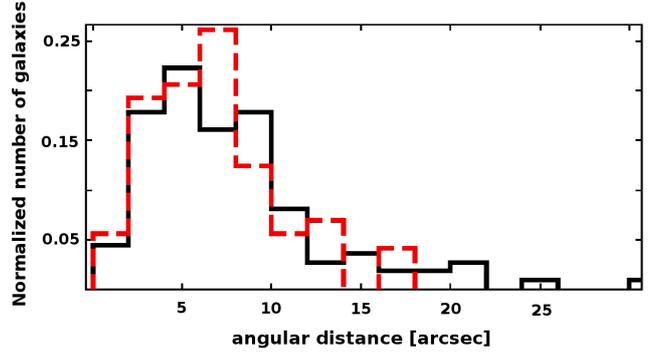}}
	\caption[]{The distribution of angular distance between AKARI ADF--S sources and optical counterparts. 
	Dashed red line histogram corresponds to the final 73 galaxies with known spectroscopic redshift,  solid black line histogram  to the final 113 galaxies with photometric redshift.}
	\label{angular_distance}
\end{figure}

\begin{figure*}[t]\centering
 	\resizebox{1\hsize}{!}{\includegraphics{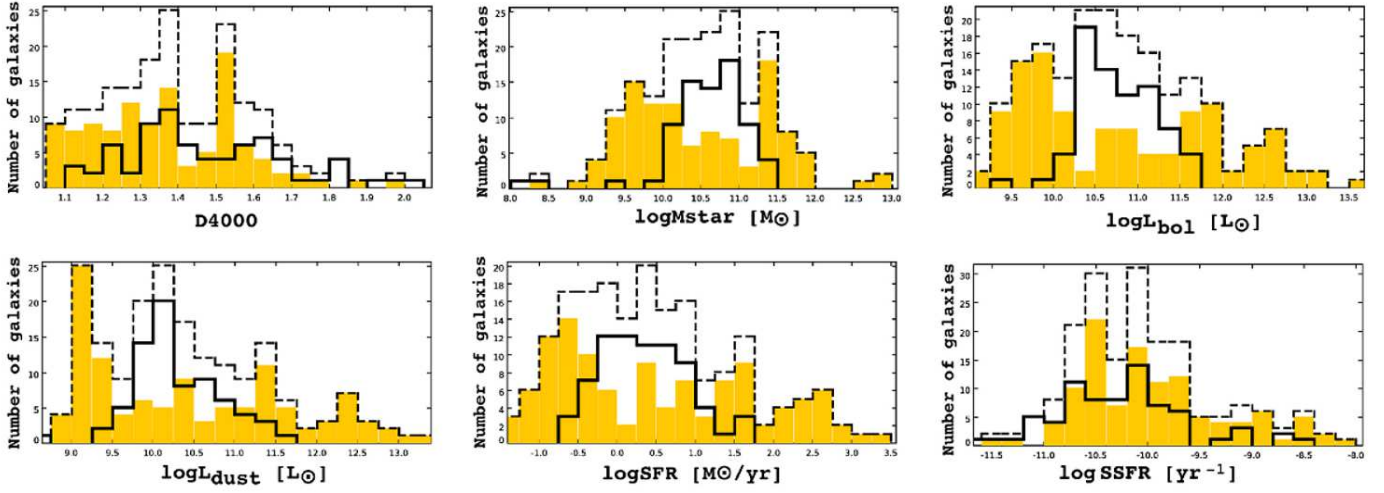}}
         \caption{Distribution of the Bayesian estimates of the output parameters: D4000, $\mathrm{M_{star}}$, $\mathrm{L_{bol}}$, $\mathrm{L_{dust}}$,  $\mathrm{logSFR}$, and  $\mathrm{logSSFR}$. 
Dotted black  line corresponds to all 186 galaxies with $\rm{\chi^2_{min}<4}$. 
Open histogram corresponds to the sample of galaxies with spectroscopic redshifts ($z_\mathrm{spec}$), and filled histogram  to galaxies with estimated photometric redshifts  ($z_\mathrm{CIGALE}$).}
\label{comparison_param_both_samples}     
  \end{figure*}

Our galaxies are almost uniformly spread across the ADF--S field.
For the subsequent analysis, we split them into two main groups: galaxies with spectroscopic $\rm{z_{spec}}$, and those with only photometric $\rm{z_{CIGALE}}$ redshifts. 
The detailed information is shown in Table~\ref{groups}, and the main parameters estimated with CIGALE for both groups are plotted in Figure~\ref{comparison_param_both_samples}. 
{ The distribution of angular distance  between the AKARI ADF--S source and the optical counterparts for the final sample of galaxies is shown in Fig.~\ref{angular_distance}.}
In Table~\ref{tab_comp_param} we present the comparison of mean values of main parameters for all successfully fitting galaxies with $\mathrm{z_{spec}}$ and $\mathrm{z_{CIGALE}}$.

\begin{table}
\caption[]{Two groups of galaxies used for final analysis.}
\label{groups}
\begin{tabular}{l l|l|l}
\multicolumn{2}{c|}{Type of }      & No. of    & No of galaxies    \\
\multicolumn{2}{c|}{redshift}      & galaxies & with SEDs$^{*}$\\ \hline \hline
\multicolumn{2}{l|}{spectroscopic} & 129      & 73 	       \\
\multicolumn{2}{l|}{$\rm{z_{CIGALE}}$}   & 127  & 113          \\ \hline
\multicolumn{2}{c|}{$\Sigma$ }     & 256     & 186           \\ \hline
\multicolumn{3}{l}{\scriptsize{* with $\rm{\chi^2_{reduced}}$ value of SEDs fitting lower than 4}}\\
\end{tabular} 
\end{table}

\begin{table}[ht]
\caption[]{The comparison of mean values of main parameters for all successfully fitting galaxies with spectroscopic redshift ($\mathrm{SAMPLE\mbox{ }z_{spec}}$) and for the sample with estimated redshift ($\mathrm{SAMPLE\mbox{ }z_{CIGALE}}$). }
\label{tab_comp_param}
\begin{tabular}{l | l | l   }
parameter &{$\mathrm{SAMPLE\mbox{ }z_{spec}}$} & {$\mathrm{SAMPLE\mbox{ }z_{CIGALE}}$} \\ \hline \hline
$\mathrm{logM_{star}}$ [$\mathrm{M_\odot}$] 		& 10.59 $\pm$ 0.14    	& 10.46 $\pm$ 0.20 \\
$\mathrm{logL_{bol}}$  [$\mathrm{L_\odot}$] 		& 10.74 $\pm$ 0.05    	& 10.81 $\pm$ 0.07 \\
$\mathrm{logL_{dust}}$ [$\mathrm{L_\odot}$] 		& 10.27 $\pm$ 0.08    	& 10.44 $\pm$ 0.08 \\
$\mathrm{logSFR}$  [$\mathrm{M_{\odot} yr^{-1}}$]       &  0.33 $\pm$ 0.18    	& 0.52 $\pm$ 0.18 \\
D4000 		  		                        & 1.47$\pm$ 0.10    	& 1.37  $\pm$ 0.92 \\
$\alpha_\mathrm{SED}$            		        & 2.05 $\pm$ 0.51    	& 1.98  $\pm$ 0.73 \\ \hline
\end{tabular} 
\end{table}

\section{Average SED}
\label{resultAVSEDS}
\begin{figure}[t]\centering
 \resizebox{1\hsize}{!}{\includegraphics{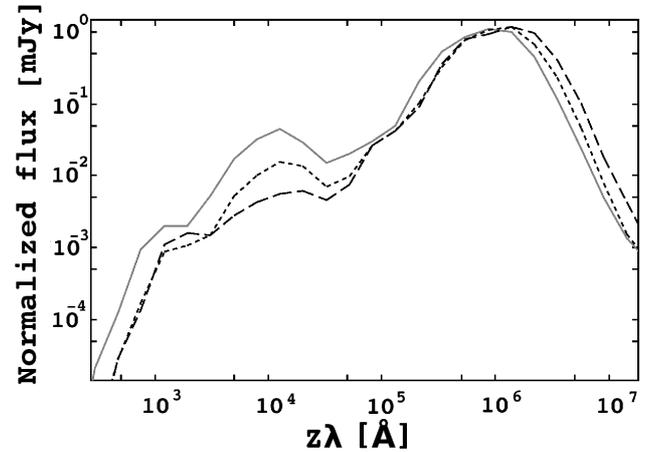}}
        \caption{The average rest-frame SED normalized at 90$\mu$m of ULIRGS (dashed line), LIRGS (dotted line) and remaining  galaxies (solid line).
        Taken from \citep{malek13}.}
     \label{Mean_sed}
\end{figure}
  
For all galaxies with $z_\mathrm{spec}$ or $z_\mathrm{CIGALE}$, we created an average SED, normalized at 90$\mu$m.
We divided our sample into three groups with respect to total infrared luminosity: ultraluminous infrared galaxies (ULIRGS, with high IR luminosity, $\mathrm{L_{TIR}>10^{12} L_{\odot}}$), luminous infrared galaxies (LIRGS, with total infrared luminosity in the range $\mathrm{10^{11}L_{\odot} < L_\mathrm{TIR}< 10^{12} L_\mathrm{\odot}}$), and the rest of the galaxies with $\mathrm{L_{TIR}<10^{11} L_{\odot}}$ .
In our sample we found 17 galaxies classified as ULIRGs (9.7\% of sources), and 31 LIRGS (16.1\% of the total number of sources) \citep{malek13}.  
All of the galaxies that belong to (U)LIRGs samples have only photometric redshift estimated by CIGALE.
The average SEDs for ULIRGs, LIRGs, and all the remaining galaxies are plotted in Figure~\ref{Mean_sed}. 

Both ULIRGS and LIRGS  contain cooler dust then the rest of our { 90$\mu$m {ADF--S} } sample with $\mathrm{L_{TIR}<10^{11} L_{\odot}}$, 
which is seen as a shift of maximum dust emission to longer wavelengths. 
The brighter the sample is in the IR range, the more shifted is the dust peak to longer wavelengths. 

{Taking the uncertainties into account, the difference in  $z\lambda_{max}$ for (U)LIRGs  is not clearly distinguishable. 
Additionally, all ULIRGs  and 74.19\% of LIRG samples have only photometric redshift information. 
Nevertheless, cold ULIRGs have been previously identified with different surveys.  
\cite{symeonidis11} found that cold ULIRGs are rather rare in our local Universe (z$<$ 0.1); however thisis result  of  the selection function rather than with the  general properties of this specific sample. 
Cold ULIRGs are detected at longer wavelengths than warm ones are.
\cite{symeonidis11} also found  that ULIRGs do not adjust to a universal luminosity--temperature correlation. 

Cold ULIRGs were previously found by
\begin{itemize}
 \item \citealp{chapman02} -  Infrared Space Observatory survey; 170$\mu$m -- selected sources at redshift $\sim$ 1; characteristic temperature equal to 30K;
 \item \citealp{chapman05} - SCUBA survey; sources detected at 850$\mu$m; temperature fitted as standard black body; characteristic temperature for this sample is equal to 36K;
 \item \citealp{kartaltepe10} - Spitzer, COSMOS survey; 70$\mu$m selected sources over the redshift range 0.01 $<$ z $<$ 3.5 with a median redshift of 0.5; they found that the COSMOS sources span the same range of colors and dust temperatures as the local sources, with apparent excess of colder sources at higher luminosities);
 \item \citealp{clements10} - SCUBA survey; 23 ULIRGs detected at 850$\mu$m; characteristic dust temperature (42K) was calculated from standard black body fitting.
 \end{itemize}
 
We conclude that our sample selection is responsible for the cooler dust temperature for (U)LIRGs samples. 
The result might be biased by uncertainty of photometric redshift estimation and the error of $z\lambda_{max}$ found for three different samples. 
}

More detailed description of (U)LIRGs samples can be found in \cite{malek13}. 
In Table~\ref{ULIRGS}, we summarize the main physical parameters for (U)LIRGS and the rest of our sample.

\begin{table}[h]
\caption[]{The main physical parameters (mean redshift, the wavelength for the maximum value of dust distribution, ratio between bolometric and IR luminosity, and the SSFR)  for ULIRGs, LIRGs, and galaxies with total infrared luminosity  $\mathrm{L_{TIR}<10^{11} L_{\odot}}$. }
\label{ULIRGS}
\begin{tabular}{l | l | l |l  }
 & ULIRGs & LIRGs &  $\mathrm{L_{TIR}<10^{11} L_{\odot}}$ \\ \hline \hline
 $z$ &0.55$\pm$0.21& 0.20$\pm$0.06& 0.05$\pm$0.03\\
$z\lambda_{max}$ [$10^{6}$\AA{}] &  1.49$\pm$0.56 & 1.25$\pm$0.63 & 0.93$\pm$0.35 \\
{\bf $\mathrm{L_{TIR}/L_{bol} } $}&0.73$\pm$0.16&0.55$\pm$0.16&0.39$\pm$0.22\\ 
logSSFR [$yr^{-1}$]& -9.00$\pm$0.55& -9.68$\pm$0.59&-10.28$\pm$0.57\\ \hline
\end{tabular} 
\end{table}

\section{Parameters obtained from galaxy SED fits}
\label{moresection}
We calculated the total dust luminosity ($L_\mathrm{TIR}$)  as the integral of spectra obtained from CIGALE in the range from 8(1+z)$\mu$m to 1(1+z) mm. 
The obtained parameter is very tightly correlated with the dust luminosity $\mathrm{L_{dust}}$, an output value from CIGALE, but the correlation is not 1:1. 
We have found that the slope of the linear correlation between logarithmic values of  $\mathrm{L_{TIR}}$ and $\mathrm{L_{dust}}$ luminosity is 1.013, with y--intercept equal to -0.136.
The Pearson product-moment correlation coefficient between  $\mathrm{L_{TIR}}$ and $\mathrm{L_{dust}}$ is equal to 0.997.
We also calculated, as an integral value from spectra, the UV luminosity of our sample ($\mathrm{L_{UV}}$), from  wavelengths 1\,480(1+z)\AA{} to 1\,520(1+z)\AA{}.

The relatively high ratio $\mathrm{L_{TIR}/L_{UV}}$ in our sample might be explained by the fact that the galaxies are selected to be very bright in the FIR band. 
The ratio of the $L_\mathrm{TIR}$ and $\mathrm{L_{UV}}$ is well correlated with dust attenuation for young stellar population. As mentioned above, this correlation is expected from the assumptions of the model. 
The correlation between $\mathrm{L_{TIR}/L_{UV}}$ and the V-band attenuation of young stellar populations can be described by a linear function:
\begin{equation}
 \mathrm{A_{V,ySP}= (0.28 \pm 0.02)log\frac{L_{TIR}}{L_{UV}}-(0.18 \pm 0.08)}.
 \label{eq1}
\end{equation}

Comparison of the total infrared luminosity with the depth of D4000 shows another correlation. 
Based on our sample, we have found the following relation between $\mathrm{L_{TIR}}$ and D4000:
\begin{equation}
 \mathrm{logL_{TIR}=-(2.69 \pm 0.32)D4000+(14.17 \pm 0.45)}.
 \label{eq2}
\end{equation}
Those galaxies which are brightest in the infrared tend to have a weak ($<$1.5) D4000.
Thus, the age of the stellar population is related to the total luminosity of dust.
In our sample, galaxies with the highest total infrared luminosity and the weakest 4000\AA{} break are at the highest redshifts.
Additionally, the nonlinear increase in dust luminosity is visible. 
This can be explained by the fact that more luminous disk galaxies have relatively more dust-enshrouded stars \citep{spinoglio95}.
\cite{spinoglio95}, used IRAS 12$\mu$m band data and the near-infrared J, H, and K bands, and in some cases the L band, to determine multiwavelength energy distribution and bolometric luminosities. 
They found that far-infrared luminosities rise faster than linearly with bolometric luminosity. 
This is also seen in our sample, but we should also take into account that galaxies very bright in $\mathrm{L_{TIR}}$ also have high photometric redshift.

\begin{figure}[t]
	\resizebox{1\hsize}{!}{\includegraphics{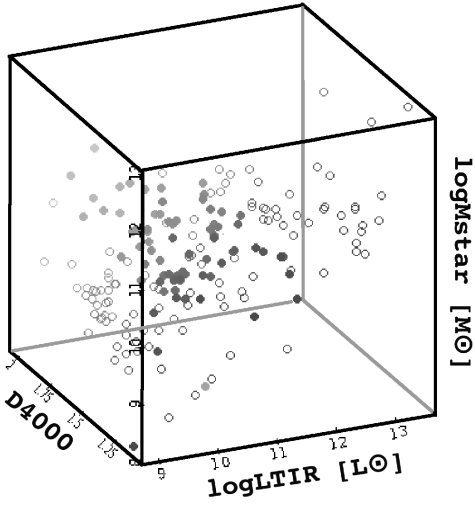}}
	\caption[]{The $\mathrm{L_{TIR}}$, D4000, and $\mathrm{M_{star}}$  relation.
Full circles correspond to galaxies with spectroscopic redshifts. 
Galaxies with estimated redshifts are shown as open circles.}
	\label{LTIR_vs_D400_vs_Mstar}
\end{figure}

\begin{figure}[t]
	\resizebox{1\hsize}{!}{\includegraphics{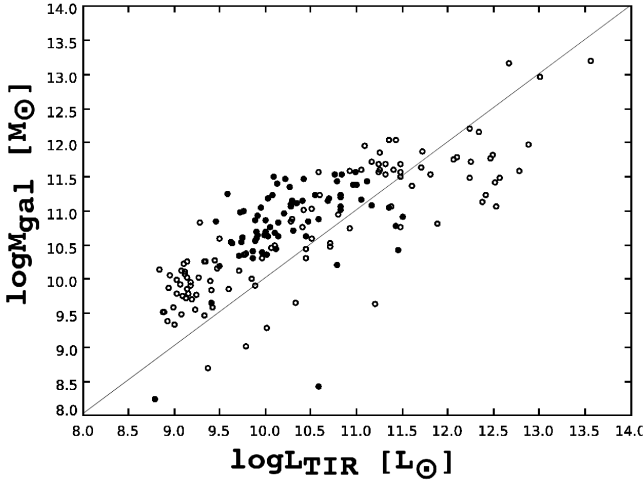}}
	\caption[]{Relation between ratio of total infrared luminosity and total mass of galaxy.
Full circles correspond to galaxies with spectroscopic redshifts. 
Galaxies with estimated redshifts are shown by open circles. }
	\label{LTIR_MGAL_MM}
\end{figure}

\begin{figure}[t]
	\centering
	\resizebox{1\hsize}{!}{\includegraphics{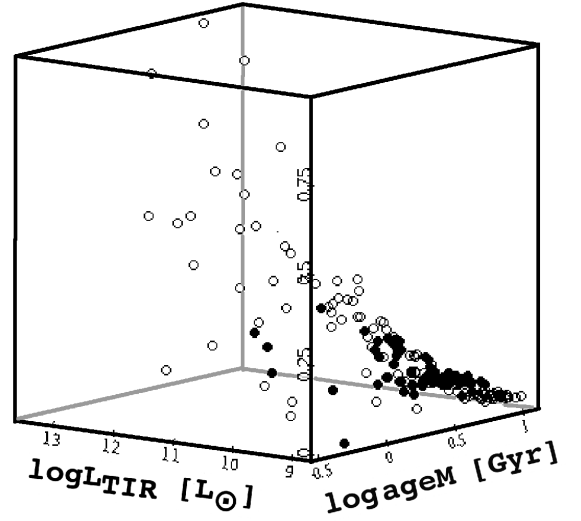}}
	\caption[]{The relation between $\mathrm{logL_{TIR}}$, and  mass weight age in function of redshift.
Full circles correspond to the galaxies with spectroscopic redshifts. 
Galaxies with estimated redshifts are marked as open circles.}
	\label{LTIR_vs_logageM}
\end{figure}

The $\rm{age_M-L_{TIR}}$ relation is plotted in Figure~\ref{LTIR_vs_logageM}.
Galaxies with redshift higher than 0.10 show a decline of mass-weighted age ($\rm{age_M}$) with increasing dust luminosity.
The slope of this decrease is 0.20.
The mass-weighted age for the rest of the sample seems to be  constant at  a level of 1 Gyr.

\section{Conclusions}
\label{conclsection}

We built a large multiwavelength catalog of 519 ADF--S galaxies brighter than 0.0301~Jy in the WIDE--S (90$\mu$m) AKARI band. 
For 129 of them, spectroscopic redshifts were obtained from public databases and \citet{sedgwick11} surveys. 

The CIGALE \citep{noll09} program for fitting spectral energy distribution, was used first to estimate photometric redshifts, and then to fit SED models to our sample.
Although CIGALE was not created to estimate $z_\mathrm{phot}$, it is more efficient than the standard code  (i.e., Le PHARE), which is based mainly on  optical and NIR data.

Spectral energy distributions were fitted for all 256 galaxies with known redshift (spectroscopic or photometric estimated by CIGALE), and for 186 galaxies the  $\rm{\chi^2_{reduced}}$ parameter for the fit was lower than 4.
The reliability of the retrieved parameters, for galaxies with known spectroscopic redshifts, was checked using a mock cataloge of artificial galaxies.
Finally 186 galaxies (73 with spectroscopic and 113 with photometric redshifts) were used for detailed  analysis. 

We conclude that FIR sources detected in the ADF--S field and selected in the 90$\mu$m band are mostly nearby galaxies (with a few cases of redshifts up to z$sim$1). 
This result is consistent with spectroscopic redshift distribution, with a mean value equal to 0.07, and also the distribution of estimated photometric redshifts (median value 0.06), and also with \cite{white12}. 

Successful SED model fits suggest that our sample galaxies are massive and IR-bright  (25\% are (U)LIRGs).
Galaxies with known spectroscopic redshifts are usually brighter in the optical range, and also with somewhat lower redshifts. 
Although our fits do not tightly constrain the  \cite{dale02} model for IR emission from dust, they indicate that our sample consists of normal
spiral{/lenticular} galaxies, with relatively active star formation.  
At the same time, the inferred distribution of the D4000 parameter suggests that most of the 186 galaxies are young and dust-rich.
The large stellar masses we find also indicate a high gas-phase metallicity, with a median value of 8.19. 

Our sample galaxies have rather high intrinsic star formation rates (with a median value  equal to 1.96 and 2.56, for galaxies with known and estimated redshift, respectively). 
The ratio of the current SFR to the total stellar mass--the SSFR--also suggests that our sample is actively star-forming, with
a characteristic timescale of less than 1 Gyr for the more 130 galaxies (70\%).

We found  a linear correlation between V-band attenuation and the ratio of $\rm{{L_\mathrm{TIR}}/{L_\mathrm{UV}}}$, and total IR luminosity and D4000. 
These relations may result from  the specific selection of our sample, down to a fixed IR flux.

Average SEDs for ULIRGs, LIRGs,  and the remaining galaxies in our sample, normalized at 90$\mu$m,  were created. 
Significant shifts of the spectral peak dust distribution, and different ratios between luminosities in the optical and IR spectra were noticed for these three samples.

\begin{acknowledgements}
This work is based on observation with AKARI, a JAXA project with the participation of ESA. 
This research has made use of the SIMBAD database, operated at CDS, Strasbourg, France, the NASA/IPAC Extragalactic Database (NED) which is operated by the Jet Propulsion Laboratory, California Institute of Technology, under contract with the National Aeronautics and Space Administration.
This research has made use of the NASA/ IPAC Infrared Science Archive, which is operated by the Jet Propulsion Laboratory, California Institute of Technology, 
under contract with the National Aeronautics and Space Administration.
KM and AP were financed by the research grants of the Polish Ministry of Science N N203 512938, and UMO-2012/07/B/ST9/04425.
Our research were conducted in the scope of the HECOLS International Associated Laboratory.
MM acknowledges support from NASA grants NNX08AU59G and NNX09AM45G for analysis of GALEX data in the Akari Deep Fields. 
TTT has been supported by the Grant-in- Aid for the Scientific Research Fund (23340046, and 24111707). 
TTT and KM have received support from  the Global COE Program Request for Fundamental Principles in the Universe: from Particles to the Solar System and the Cosmos, Strategic Young Researchers Overseas Visits Program for Accelerating Brain Circulation  commissioned by the
Ministry of Education, Culture, Sports, Science and Technology (MEXT) of Japan.
\end{acknowledgements}

\bibliographystyle{aa} 



\newpage
\onecolumn{
\clearpage
\label{tableallzcigale}
\begin{longtable}{c|l|l|l|l}
\caption[]{Photometric redshifts of 113 galaxies estimated by CIGALE. 
All listed sources have at least six photometric measurements in the whole spectral range, and a  $\chi^{2}$ value of SED fitting lower than 4. } \tabularnewline
$ID_\mathrm{ADF-S}$ & ra [deg] & dec[deg] & counterpart & $z_\mathrm{CIGALE}$ \\
\hline \hline
\endfirsthead
\caption{continued.}\tabularnewline
$ID_\mathrm{ADF--S}$ & ra [deg] & dec[deg] & counterpart & $z_\mathrm{CIGALE}$ \\
\hline \hline
\endhead
\hline \hline
\endfoot
12 &  69.10183 & -54.41481 & 2MASX  J04362406-5424552 & 0.06 \\ 
13 &  68.59812 & -54.69248 & 2MASX  J04342317-5441331 & 0.20 \\ 
14 &  69.55033 & -54.37286 & 2MASX  J04381194-5422292 & 0.22 \\ 
15 &  70.42883 & -52.43447 & 2MASX  J04414298-5226042 & 0.02 \\ 
19 &  73.29128 & -52.90457 & 2MASX  J04530951-5254202 & 0.02 \\ 
20 &  66.76948 & -54.26188 & 2MASX  J04270401-5415463 & 0.04 \\ 
22 &  68.96996 & -54.40260 & 2MASX  J04355267-5424143 & 0.04 \\ 
24 &  68.87662 & -53.31383 & 2MASX  J04352989-5318515 & 0.16 \\ 
26 &  67.11937 & -55.02170 & APMUKS(BJ) B042722.82-550755.3 & 0.38 \\ 
31 &  67.13429 & -53.99541 & 2MASX  J04283256-5359474 & 0.02 \\ 
41 &  68.14438 & -53.79695 & 2MASX  J04323435-5347510 & 0.06 \\ 
46 &  69.86474 & -52.54273 & 2MASX  J04392756-5232345 & 0.20 \\ 
47 &  67.58768 & -53.70834 & APMUKS(BJ) B042911.18-534903.9 & 0.54 \\ 
49 &  66.98278 & -53.80051 & APMUKS(BJ) B042646.40-535442.5  & 0.06 \\ 
51 &  69.72732 & -52.86293 & ESO 157-46 & 0.02 \\ 
53 &  67.34102 & -54.07068 & 2MASX  J04292162-5404154 & 0.02 \\ 
55 &  75.64549 & -53.19288 & 2MASX  J05023480-5311321 & 0.02 \\ 
59 &  68.36949 & -54.55663 & APMUKS(BJ) B043219.97-543942.2 & 0.50 \\ 
60 &  67.22106 & -53.27556 & 2MASX  J04285263-5316337 & 0.14 \\ 
61 &  75.09628 & -52.64538 & 2MASX  J05002294-5238446 & 0.98 \\ 
63 &  68.03277 & -53.15993 & 2MASX  J04320736-5309369 & 0.38 \\ 
67 &  69.86035 & -53.04714 & ESO 157-48 & 0.04 \\ 
68 &  75.14589 & -53.31892 & 2MASX  J05003472-5319056 & 0.02 \\ 
69 &  65.97959 & -54.00145 & 2MASX  J04235487-5400082 & 0.06 \\ 
74 &  66.58741 & -53.67844 & APMUKS(BJ) B042510.28-534732.0 & 0.46 \\ 
78 &  67.20383 & -53.64476 & APMBGC 157-030-071 & 0.06 \\ 
82 &  67.92904 & -53.86149 & APMUKS(BJ) B043035.15-535810.7 & 0.02 \\ 
83 &  73.24695 & -52.14472 & APMUKS(BJ) B045146.99-521333.6 & 0.32 \\ 
93 &  74.76365 & -53.18383 & 2MASX  J04590078-5310503 & 0.20 \\ 
96 &  66.70094 & -54.14094 & APMUKS(BJ) B042539.30-541513.0 & 0.88 \\ 
101 &  74.45489 & -52.67251 & 2MASX  J04574983-524036 & 0.04 \\ 
102 &  69.81444 & -52.95298 & APMUKS(BJ) B043804.54-530301.8 & 0.52 \\ 
104 &  72.94480 & -51.67736 & 2MASX  J04514705-514036 & 0.02 \\ 
111 &  69.53889 & -54.01321 & 2MASX  J04380870-540050 & 0.06 \\ 
113 &  67.71205 & -53.79554 & 2MASX  J04305049-534749 & 0.02 \\ 
114 &  67.73132 & -55.36489 & 2MASX  J04305608-552155 & 0.02 \\ 
117 &  75.17566 & -52.73285 & 2MASX  J05004165-5243575 & 0.10 \\ 
119 &  71.29812 & -52.32486 & 2MASX  J04451081-5219279 & 0.10 \\ 
122 &  71.52372 & -53.83169 & 2MASX  J04460539-534955 & 0.02 \\ 
123 &  73.27410 & -51.73525 & 2MASX  J04530553-514401 & 0.02 \\ 
126 &  68.18842 & -54.26613 & 2MASX  J04324535-541601 & 0.02 \\ 
139 &  67.89508 & -54.07494 & APMUKS(BJ) B043023.58-541057.6 & 0.18 \\ 
142 &  72.95174 & -51.64787 & APMUKS(BJ) B045035.16-514343.1 & 0.04 \\ 
143 &  71.79445 & -53.76829 & 2MASX  J04471047-534604 & 0.02 \\ 
146 &  74.82765 & -52.85153 & 2MASX  J04591744-5251052 & 0.20 \\ 
147 &  72.03197 & -53.42830 & 2MASX  J04480825-532539 & 0.02 \\ 
148 &  73.22942 & -51.59712 & 2MASX  J04525430-513542 & 0.24 \\ 
151 &  68.03200 & -53.91074 & APMUKS(BJ) B043059.13-540056.2 & 0.54 \\ 
154 &  74.13599 & -52.93473 & 2MASX  J04563217-525605 & 0.24 \\ 
159 &  71.79180 & -53.76501 & 2MASX  J04471047-534604 & 0.02 \\ 
165 &  66.71423 & -54.40252 & 2MASX  J04265073-542415 & 0.36 \\ 
170 &  74.41248 & -52.23736 & APMUKS(BJ) B045627.32-521846.4 & 0.04 \\ 
173 &  74.43239 & -52.33703 & 2MASX  J04574286-522009 & 0.04 \\ 
175 &  74.60095 & -53.46246 & 2MASX  J04582396-5327481 & 0.20 \\ 
177 &  74.45272 & -52.56456 & 2MASX  J04574760-523355 & 0.02 \\ 
179 &  69.33923 & -53.09289 & 2MASX  J04372100-530537 & 0.02 \\ 
190 &  74.45093 & -52.56502 & 2MASX  J04574760-523355 & 0.02 \\ 
191 &  72.29425 & -53.77261 & 2MASX  J04490850-534630 & 0.14 \\ 
192 &  72.19001 & -52.51645 & 2MASX  J04484409-523054 & 0.04 \\ 
198 &  73.97142 & -51.50573 & APMUKS(BJ) B045439.83-513456.4 & 0.60 \\ 
203 &  72.97690 & -52.26350 & APMUKS(BJ) B045044.73-522017.4 & 0.30 \\ 
205 &  70.38938 & -52.36417 & 2MASX  J04413259-522148 & 0.04 \\ 
206 &  68.16271 & -54.42572 & APMUKS(BJ) B043131.27-543152.8 & 0.92 \\ 
210 &  70.65664 & -52.53704 & 2MASX  J04423732-523210 & 0.02 \\ 
212 &  71.91896 & -52.02849 & APMUKS(BJ) B044627.85-520656.1 & 0.74 \\ 
213 &  68.56086 & -53.93207 & 2MASX  J04341529-535605 & 0.02 \\ 
215 &  70.68093 & -52.28554 & 2MASX  J04424423-5216587 & 0.20 \\ 
219 &  75.15276 & -53.03395 & 2MASX  J05003677-530153 & 0.02 \\ 
222 &  70.20829 & -54.21960 & APMUKS(BJ) B043943.07-541847.8 & 0.62 \\ 
225 &  72.44701 & -52.96709 & 2MASX  J04494618-525808 & 0.02 \\ 
234 &  73.29361 & -52.40390 & APMUKS(BJ) B045159.43-522906.2 & 0.24 \\ 
247 &  73.52169 & -52.70868 & 2MASX  J04540432-524232 & 0.02 \\ 
252 &  74.89770 & -52.24227 & 2MASX  J04593486-521433 & 0.16 \\ 
260 &  68.34128 & -54.65844 & APMUKS(BJ) B043214.79-544545.0 & 0.46 \\ 
265 &  70.15438 & -52.68969 & 2MASX  J04403699-524126 & 0.18 \\ 
\label{allzcigale}
\end{longtable}
}

\end{document}